\documentclass[pra,aps,superscriptaddress,twocolumn,floatfix,notitlepage,longbibliography]{revtex4-2}

\usepackage{amsmath}
\usepackage{amsfonts}
\usepackage{amssymb}
\usepackage{graphicx}
\usepackage{color, soul}
\usepackage{xcolor}
\usepackage{xspace}
\usepackage{physics}
\usepackage{float}

\newcommand{\ud}[1]{{#1^{\dagger}}}
\newcommand{\av}[1]{{\langle #1 \rangle}}

\begin{document}

\title{Spectral correlations of dynamical Resonance Fluorescence}

\author{Santiago Bermúdez-Feijóo}
\thanks{These two authors contributed equally.}
\affiliation{Institute for Photonic Quantum Systems (PhoQS), Center for Optoelectronics and Photonics
Paderborn (CeOPP), and Department of Physics, Paderborn University, 33098 Paderborn, Germany}

\author{Eduardo Zubizarreta Casalengua} 
\thanks{These two authors contributed equally.}
\affiliation{Walter Schottky Institute, School of Computation, Information and Technology and MCQST, Technische Universität München, 85748 Garching, Germany}

\author{Kai Müller}
\affiliation{Walter Schottky Institute, School of Computation, Information and Technology and MCQST, Technische Universität München, 85748 Garching, Germany}

\author{Klaus D. Jöns}
\affiliation{Institute for Photonic Quantum Systems (PhoQS), Center for Optoelectronics and Photonics
Paderborn (CeOPP), and Department of Physics, Paderborn University, 33098 Paderborn, Germany}


\date{\today}

\begin{abstract}
Frequency-filtered photon correlations have been proven to be extremely useful in grasping how the detection process alters photon statistics. Harnessing the spectral correlations also permits refinement of the emission and unraveling of previously hidden strong correlations in a plethora of quantum-optical systems under continuous-wave excitation. In this work, we investigate such correlations for time-dependent excitation and develop a methodology to compute efficiently time-integrated correlations, which are at the heart of the photon-counting theory, and subsequently apply it to analyze the photon emission of pulsed systems. By combining this formalism with the \emph{sensor method}—which facilitates frequency-resolved correlations—we demonstrate how spectral filtering enhances single-photon purity and suppresses multi-photon noise in time-bin-encoded quantum states. Specifically, filtering the central spectral peak of a dynamically driven two-level system boosts temporal coherence and improves the fidelity of time-bin entanglement preparation, even under conditions favoring multi-photon emission. These results establish spectral filtering as a critical tool for tailoring photon statistics in pulsed quantum light sources.
\end{abstract}

\maketitle

\section{Introduction}
The interaction between two-level quantum systems (TLS) and resonant electromagnetic fields is fundamental in quantum optics, deepening our understanding of light-matter interactions and developing photonic quantum technologies. When a resonant field continuously drives a TLS, resonance fluorescence produces the Mollow triplet \cite{mollow_power_1969, flagg_resonantly_2009} spectrum at strong driving. This behavior, observed as a central peak flanked by symmetric sidebands in the emission spectrum, is further enriched when examining photon correlations between these spectral components, revealing that the perfect antibunched nature of the driven TLS extends into a wide range of photon-pair behaviors, that go from antibunching to strong bunching \cite{lopez_carreno_photon_2017, peiris_two-color_2015, gonzalez-tudela_two-photon_2013, nieves_third-order_2020}. Such frequency-filtered correlations \cite{del_valle_theory_2012} have shown promise for single-photon emitters and other quantum technologies, enabling finely tunable photon sources with high degrees of control over photon statistics \cite{kim_unlocking_2024}, as well as demonstrating entanglement over the symmetric sidebands of the Mollow triplet spectrum \cite{lopez_carreno_entanglement_2024}.
\\\\Recent advancements in the analysis of fluorescence spectra have extended from continuous-wave (CW) to pulsed excitation schemes, which have opened new avenues in controlling two-level systems. Under Gaussian pulses, dynamic resonance fluorescence emerges, exhibiting new emission patterns with sidebands that depend on pulse intensity and duration \cite{moelbjerg_resonance_2012}. This prediction was recently experimentally confirmed in both Semiconductor QDs \cite{boos_signatures_2024} and in Solid-State cavity-QED systems \cite{liu_dynamic_2024}. The dynamic driving allows for an intricate interplay between photon statistics, temporal correlations, and spectral behavior. Theoretical \cite{senellart_high-performance_2017} and experimental work \cite{santori_triggered_2001, santori_indistinguishable_2002} has demonstrated that pulsed excitation not only produces coherent single-photon emission \cite{he_-demand_2013} but also leads to exotic photon states and multi-photon bundles, paving the way for advanced quantum light sources \cite{loredo_generation_2019, vajner_towards_2024}.
\\\\Despite significant progress in understanding pulsed-driven two-level systems, the role of frequency-filtered correlations remains unexplored. Prior studies have shown that photon statistics can oscillate between bunching and anti-bunching depending on the pulse area \cite{fischer_signatures_2017, konthasinghe_correlations_2015}, yet, these analyses often neglect the impact of frequency filtering the signal. Recent work by López Carreño \cite{carreno_cascaded_2024} demonstrated how pulsed excitation enables cascaded single-photon emission, while Redivo Cardoso \textit{et al.} \cite{cardoso_impact_2025}  explored the interplay of temporal correlations and decoherences in the biexciton-exciton cascades. Given the importance of photon correlations in quantum applications \cite{kimble_quantum_2008}, a deeper investigation into frequency-dependent photon correlations in the time-dependent regime is needed.
\\\\In this work, we extend the understanding of time-dependent photon correlations by exploring frequency-filtered two-photon correlations in a two-level system driven by Gaussian pulses. We show that, under pulsed excitation, the correlations between photons emitted from different spectral regions exhibit a rich and controllable structure: frequency resolution decouples photon statistics from the pulse area alone, enabling transitions between antibunching and bunching depending on the selected filter frequencies and bandwidths.
 Furthermore, we demonstrate that spectral filtering enables precise control over time-bin purities, suppressing multi-photon events while enhancing single-photon fidelity. By filtering the central spectral peak, temporal coherence is enhanced, restoring single-photon dominance even under conditions favoring multi-photon emission (e.g., at even pulse areas). This study builds a versatile framework for tailoring photon statistics in pulsed systems, directly addressing challenges in time-bin-encoded quantum state preparation and advancing applications in quantum communication and computing.

\section{Photon correlation functions}
\subsection{Ordered and symmetric photon correlation functions}
Given that in this work we will be dealing with the calculations of photon correlations of different frequency modes and times, it is necessary to make a clear distinction between \textit{ordered} and \textit{symmetric} photon correlations. In general, the $n$-photon correlations are defined as
\begin{equation}
\resizebox{0.5\textwidth}{!}{$    \mathcal{G}^{(n)}_{a_1 a_2 ... a_n } (t_1, t_2, ..., t_n) =
    \left\langle \mathcal{T}_+ [\prod_{k = 1}^n \ud{a_k} (t_k) ] 
    \mathcal{T}_- [\prod_{k' = 1}^n a_{k'} (t_{k'}) ]
    \right\rangle, $}
\end{equation}
where $\mathcal{T}_{\pm}$ orders the operators from right to left (left to right) for rising times, and $a_k$ can be any annihilation operator we may consider, corresponding to the same or different modes.
These correlation functions are, by definition, \textit{symmetric} functions. This means that every exchange of any pair of times and operators ($t_k \leftrightarrow t_{k'}$
and $a_k \leftrightarrow a_{k'}$) would leave the functions unchanged. 
For instance, the two-photon autocorrelation function $\mathcal{G}^{(2)}_{aa} (t_1,t_2)$
is symmetric when we swap $t_1$ and $t_2$, i.e., $\mathcal{G}^{(2)}_{aa} (t_2,t_1) 
= \mathcal{G}^{(2)}_{aa} (t_1,t_2)$.
\\\\
On the other hand, we need to introduce the \textit{ordered} correlation functions,
where the times $t_k$ follow a strict ordering that, for convenience, we assume to be as follows: $t_1 < t_2 < \hdots < t_n$. In such a case, we use the notation 
$\mathcal{G}^{(n)}_{a_1 \rightarrow a_2 \hdots \rightarrow a_n } (t_1, t_2, \hdots, t_n)$.
Then, the cross-correlation $\mathcal{G}^{(2)}_{ab} (t_1,t_2)$ is simply written in terms of the ordered correlation functions as
\begin{equation}
 \mathcal{G}^{(2)}_{ab} (t_1,t_2) = \theta(t_2 - t_1) \mathcal{G}^{(2)}_{a \rightarrow b}
 (t_1, t_2)  + \theta(t_1 - t_2) \mathcal{G}^{(2)}_{b \rightarrow a} (t_2,t_1) \,,
\end{equation}
where $\theta(x)$ is the Heaviside function, which returns 1 only if $x > 0$
and is zero otherwise. For completeness, we also define the correlation functions
in terms of the delays $\tau_k \equiv t_{k+1} - t_k$ between successive times and 
time $t$ (recasting $t_1$ as $t$).
The ordered correlation function is then expressed as
\begin{align}
    \small
    &G^{(n)}_{a_1 a_2 \hdots a_n } (t, \tau_1, \hdots, \tau_{n-1}) = \langle 
    \ud{a_1} (t) \ud{a_2} (t+\tau_1)\hdots\\\nonumber  
    &\hdots \ud{a_n} 
    a_n (t+\tau_1 + \hdots + \tau_{n-1}) \hdots a_2 (t+\tau_1) a_1 (t)
    \rangle \,,
\end{align}
which is more convenient when it comes to solving the $n$-times dynamics
through the \textit{Quantum Regression Theorem} (QRT) \cite{carmichael_open_1993, breuer_theory_2007}. We can, however, naturally connect
both symmetric and ordered correlation functions. Selecting, for instance $n=2$, gives the following auto-correlation function, 
\begin{align*}
 \mathcal{G}^{(2)}_{aa} (t_1,t_2) &=
 \theta(t_2 -t_1) G_{aa}^{(2)} (t_1, t_2 - t_1) \\\nonumber
 &+\theta(t_1 - t_2)
  G_{aa}^{(2)} (t_2, t_1 - t_2) \,,
\end{align*}
whereas the cross-correlations are
\begin{align*}
 \mathcal{G}^{(2)}_{ab} (t_1,t_2) &=
 \theta(t_2 -t_1) G_{ab}^{(2)} (t_1, t_2 - t_1) \\
 &+\theta(t_1 - t_2)
  G_{ba}^{(2)} (t_2, t_1 - t_2) \,,
\end{align*}
where
\begin{align*}
    G_{aa}^{(2)} (t,\tau) &= \av{\ud{a}(t) \ud{a}a(t+\tau) a(t)} \,, \\
    G_{ab}^{(2)} (t,\tau) &= \av{\ud{a}(t) \ud{b}b(t+\tau) a(t)} \,, \\
    G_{ba}^{(2)} (t,\tau) &= \av{\ud{b}(t) \ud{a}a(t+\tau) b(t)} \,.
\end{align*}
It is important to understand that $G_{ab}^{(2)} (t,\tau)$ and $G_{ba}^{(2)} (t,\tau)$ refer to
different processes, as the former considers the photon emission from
the mode $b$ preceded by a photon from mode $a$ ($a \rightarrow b$), while the latter accounts 
the reversed process ($b \rightarrow a$). 

\subsection{Time integrated correlation functions}
With the previous definitions, we now introduce the time integrated correlation functions
\begin{multline}
\label{eq:IntegratedGN_def}
G_{a_1 a_2 \hdots a_n }^{(N)}[0,T] = \\
\int_0^T \dots \int_0^T \mathcal{G}^{(N)}_{a_1 a_2 \hdots a_n } (t_1, t_2, \dots,t_N) \, dt_1 dt_2 \dots dt_N \,,
\end{multline}
where $T$ defines the span of the time bin, that is, from 0 to $T$.
Hereafter, we focus on the $N=2$ case, which reads
\begin{equation}
\label{eq:IntegratedG2_def}
G_{a_1 a_2}^{(2)}[0,T] = \int_0^T \int_0^T \mathcal{G}^{(2)}_{a_1 a_2} (t_1,t_2) \, dt_1 dt_2 \,,
\end{equation}
In terms of the ordered correlation functions, the integrated correlation
function reads
\begin{multline}
\label{eq:IntegratedG2_def2}
 G_{a_1 a_2}^{(2)}[0,T] =    \int_0^T \int_{0}^T \theta(t_2-t_1)\mathcal{G}^{(2)}_{a_1 \rightarrow a_2} (t_1,t_2) \, dt_1 dt_2\\
  +\int_0^T \int_{0}^T \theta(t_1-t_2) \mathcal{G}^{(2)}_{a_2 \rightarrow a_1} 
  (t_2,t_1) \, dt_1 dt_2
 \,,
\end{multline}
which after swapping $t_1 \leftrightarrow t_2$ in the second integral, we get
\begin{align}
   \label{eq:intG2_12}
   G_{a_1 a_2}^{(2)}[0,T] &=  \int_0^T \int_{0}^T \theta(t_2-t_1)\mathcal{G}^{(2)}_{a_1 \rightarrow a_2} (t_1,t_2) \, dt_1 dt_2\\\nonumber
  &+\int_0^T \int_{0}^T \theta(t_2-t_1) \mathcal{G}^{(2)}_{a_2 \rightarrow a_1} 
  (t_1,t_2) \, dt_1 dt_2 \\\nonumber
   &=G_{a_1 \rightarrow a_2}^{(2)}[0,T] + G_{a_2 \rightarrow a_1}^{(2)}[0,T] 
 \,.
\end{align}
This makes explicit the two contributions---depending
on the emission order---to the cross-correlations.
Changing the variables from $(t_1,t_2)$ to $(t,\tau)$,
we also find
\begin{equation}
G_{a_1 \rightarrow a_2}^{(2)}[0,T] =  
\int_0^T \int_{0}^{T-t} G^{(2)}_{a_1 a_2} (t,\tau) d \tau dt \,.
\end{equation}
This definition differs from the one given by Fischer \cite{fischer_signatures_2017} in the
upper limit of the $\tau$ integral. Nonetheless,
both quantities converge when $T \rightarrow \infty$, or,
more realistically, when $G^{(2)}$ is integrated over a domain big enough
to contain all the relevant features.\\\\
The reason why we chose the definition of Eq.~(5) over Fischer's \cite{fischer_signatures_2017} is that, in first place, the former is intimately connected
to Mandel's photon-counting theory~\cite{carmichael_open_1993, gardiner_quantum_2000} and represents the probability of detecting two photons, one in each mode, when $(n>2)$-photon events can be ruled out. Secondly, these two-fold (or $n$-fold in general) integrals can be efficiently computed from the equations of
motion that we will derive in subsequent sections. 
\\\\
We are investigating the time structure of the emission of (at least) two photons that can be either distinguishable or indistinguishable in frequency.
For such a purpose, we set two time bins. The time domain is then split in two: $0 < t \leq T$, which defines the first bin,
and $t > T$, to which we assigned the labels Early (E) and Late (L), respectively. If we extend the description to two modes and, thereby
two different times $(t_1, t_2)$, the first associated to the mode $a$, and
the second to $b$, the resulting space is split into four domains: $(\mathrm{EE})$, if both photons are detected in the Early bin;
$(\mathrm{LE})$, when $a$ is detected in the Late bin and $b$ in the Early one; $(\mathrm{EL})$, for the reversed process ($a$ in E and $b$ in L); and
$(\mathrm{LL})
$, whether both photons are detected in the Late bin.
To study these types of correlations, we require a more general definition of
time-integrated correlations than Eq.~\eqref{eq:IntegratedG2_def}, as the
integration domain is not $(0,T)$. Rather, we need to define the generalised
time-integrated two-photon correlation function, whose integration windows can be different: $0 < t_1< T$, and $0 < t_2< T'$. 
Without loss of generality, we assume that T<T'. We then recast the $T' = T + \tau$, that is, $\tau$ is the difference between the two integration times. Then, the correlation function reads
\begin{equation}
\label{eq:G2Int_gen}
    G_{a_1 \rightarrow a_2}^{(2)}[0,T; \tau] =
    \int_0^T \int_0^{T+\tau} \, \mathcal{G}^{(2)}_{a_1 \rightarrow a_2} (t_1,t_2) 
    \, dt_1 dt_2 \,,
\end{equation}
which will allow us to extract the correlations from different time bins (for the details, consult Appendices B and C).
\\\\
From the temporal correlations, given by $\mathcal{G}_{ab}(t_1, t_2)$, it is possible to analyze both the temporal spread of detections and the asymmetry of the emission. This information reveals details about the order of the photon emission process and the relevant time scale of the dynamics. However, directly quantifying the likelihood of emission in a specific scenario, such as determining the probability of detecting both photons within region $(\mathrm{EE})$, remains challenging. To address this, in Appendix \ref{appendix:PhotCount}, we derive explicit expressions for calculating the probabilities of detecting a specific number of photons within designated time bins. These calculations apply both to the single-mode case, as in bare resonance fluorescence (Sec. \ref{UN_RF}), and to the two-mode case, which corresponds to the scenario of frequency-filtered correlations (Sec. \ref{sec:FiltRF}). In addition, we check the results obtained using this method with Monte Carlo (MC) simulations for both the
bare \cite{molmer_monte_1993} and filtered \cite{lopez_carreno_frequency-resolved_2018} emission. We adapted the procedure outlined in the previous references to include time-dependent excitation.
However, although MC simulations do provide direct access to the photon counting statistics, they
require many trajectories to reach convergence. Thus, the computation
time is often long compared to the method we present here, which is
quantitatively much faster; however, the main drawback is that one needs to truncate the
correlations up to a certain photon number $N$. We show a comparison of both
methods in Fig.~\ref{ap:figMCvsInts} (Appendix~E), which allowed to determine
the validity of the results when the truncation photon number is $N = 2$.  
As for the bare case, the Monte Carlo simulations perfectly agree with
the result obtained using our method displayed in Fig.~1. Therefore, we do not explicitly show the
comparison here.
\section{Bare Resonance Fluorescence}\label{UN_RF}
Resonance fluorescence can be primarily modeled by a TLS driven by an external electromagnetic field. The dynamics of the interaction in the case of finite driving is given by the following rotating frame Hamiltonian,
\begin{align}\label{eq:dynTLSHam}
H_{\sigma}(t) = {\Tilde{\omega}}_{\sigma}\sigma^\dagger\sigma + \frac{\Omega(t)}{2}\left(\sigma^\dagger + \sigma\right),
\end{align}
where $\sigma = \ket{G}\bra{X}$ represents the lowering TLS operator, ${\Tilde{\omega}}_{\sigma} =\omega_\sigma - \omega_L$ is the detuning between the driving frequency and the transition frequency of the two-level system, and $\Omega(t)$ represents a Gaussian pulse envelope of the form,  
\begin{align}
\Omega(t) = \frac{\Theta}{\sqrt{2\pi}\tau_d}\exp\left(-\frac{(t-t_0)^2}{2\tau_{d}^{2}}\right),
\end{align}
where $\tau_d$ is the pulse duration, $t_0$ the pulse center, and $\Theta$ denotes the pulse area. The dissipative nature of the system is incorporated through a Lindblad master equation, which accounts for the spontaneous emission with a decay rate $\gamma_\sigma$,
\begin{align}\label{eq:2LSME}
\partial_t\rho &= i[\rho,H_\sigma]+\frac{\gamma_\sigma}{2}\mathcal{L}_{\sigma}\rho,
\end{align}
where the Lindblad superoperator is defined as $\mathcal{L}_c\rho = 2c\rho c^\dagger - c^\dagger c\rho - \rho c^\dagger c$. By solving this master equation, we can calculate correlation functions that will help us clarify the dynamics of the emission of the system for different time bins.
\begin{figure}[h!]
    \centering
\includegraphics[width=\columnwidth]{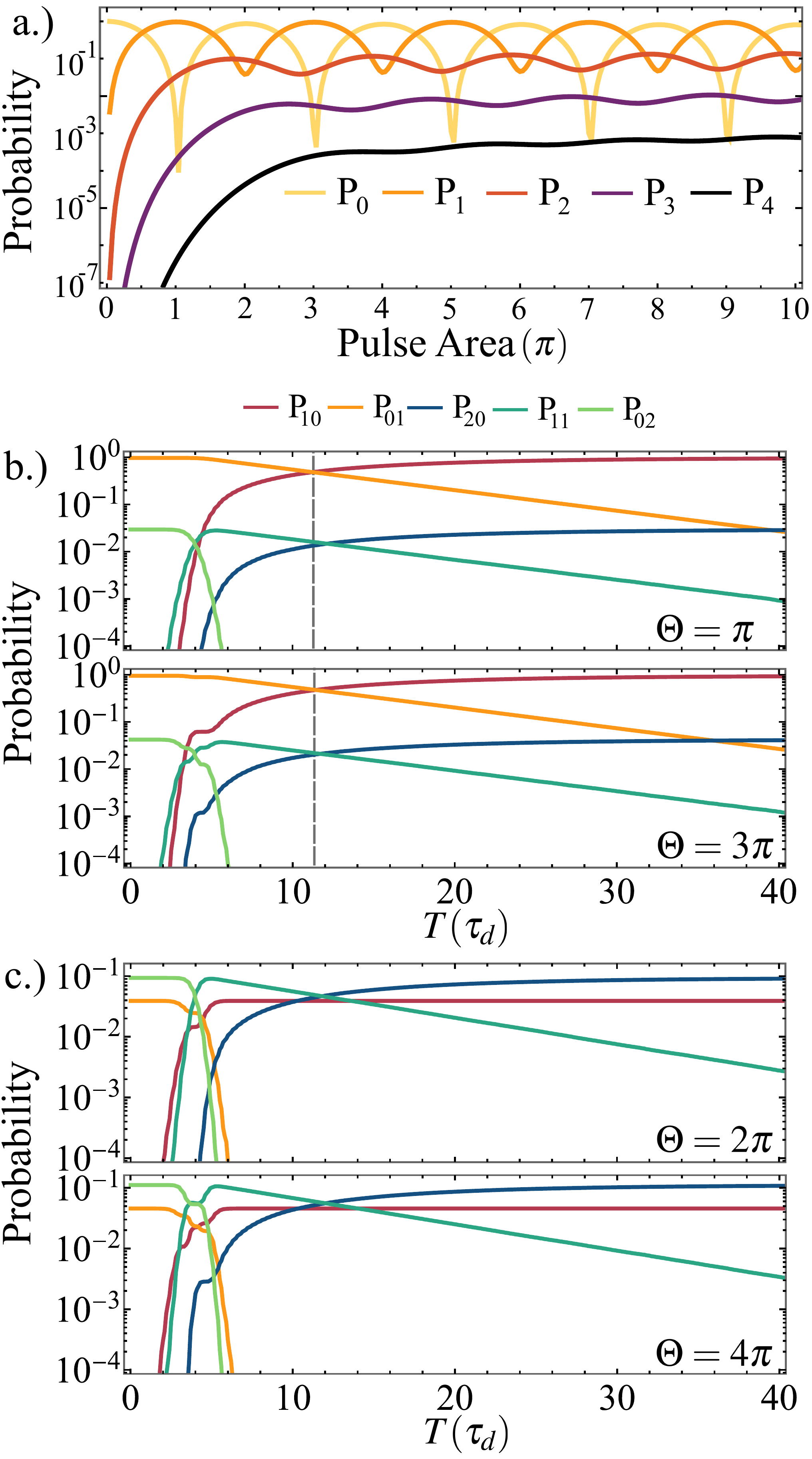}
    \caption{
    a.) Photon probabilities $\text{P}_n$ (Eq.~\eqref{eq:Pn_formula}) 
    calculated as a function of the pulse area $\Theta$.
    b, c.) Photon probabilities $\text{P}_{mn}$ (Eq.~\eqref{eq:Pmn_formula}) of finding $m, n$ photons in the early (E) and late (L) time bins, respectively, as a function of the time bin splitting parameter T, given in units of the pulse duration $\tau_d$, from the bare emission of Resonance Fluorescence, for odd (b) and even(c) pulse areas, respectively. The dashed lines shown for odd pulse areas (b) represent the value of the parameter $T$ where the timebin one photon probability is equal, i.e., $P_{01} = P_{10}$, a condition suitable for time-bin superposition state preparation. \textit{Parameters:} $\Tilde{\omega}_\sigma = 0$, $\tau_d=1/10\gamma_\sigma$, $t_0 = 4\tau_d$.}\label{fig:PhotProb2LS}
\end{figure}
\\\\Using the equations of motion provided in Appendix~B (Eqs.~\ref{eq:EqsMot_Gn}) and one-mode photon probabilities given in Appendix~C (Eq.~\ref{eq:Pn_formula}), we calculate the probabilities $\text{P}_n$ of detecting $n$ photons as a function of the pulse area ($\Theta$), for a pulse duration of $\tau_d = 1/10\gamma_\sigma$. Figure \ref{fig:PhotProb2LS}a illustrates the behavior for up to four photons over a range of pulse areas extending to $10\pi$. The results reveal a clear oscillatory pattern depending on whether $\Theta$ is even or odd. Specifically, for even pulse areas the condition  $\text{P}_2 > \text{P}_1$ holds, indicating dominance of two-photon emission, while for odd pulse areas, single-photon emission remains the most significant contribution.\\\\
Moreover, while the probabilities of three-photon and four-photon emission are on the order of $10^{-4}$ and $10^{-7}$, respectively, for $\Theta = \pi$, they increase significantly with larger pulse areas, reaching $10^{-2}$ and $10^{-3}$ when $\Theta = 10\pi$. This highlights the transition from single-photon to multiphoton emission as the dominant process, in contrast to the behavior of ideal single-photon sources. In particular, already at $\Theta = 3\pi$, the likelihood of observing three photons exceeds that of observing none. This underscores the relevance of multiphoton contributions in this regime, and suggests that the use of TLS as a single-photon source can be compromised beyond the $\pi$-pulse condition, even if the pulse area remains odd. Therefore, additional mechanisms are needed to ensure single-photon behavior at higher pulse areas.\\\\
Furthermore, in Figures \ref{fig:PhotProb2LS}b and \ref{fig:PhotProb2LS}c we analyze the temporal distribution of the emitted photons, for pulse areas ($\Theta$) between $\pi$ and $4\pi$, by splitting them into Early and Late time bins, based on the time bin parameter $T$, which is given in units of the pulse duration $\tau_d$. This corresponds to evaluating the likelihood of finding, for example, two photons within the early time bin ($\text{P}_{20}$), depending on the value of $T$. For simplicity in the visualization, we restrict our analysis to combinations involving up to two photons, as Figure \ref{fig:PhotProb2LS}a shows that these events are the most likely to occur, compared to those with three and four photons.\\\\
At the beginning of Figure \ref{fig:PhotProb2LS}b, all single and two-photon events are located within the late time bin, with the probabilities remaining constant until $T \approx 3 \tau_d$. As $T$ increases, the likelihood of photons being detected in different bins also increases. For two-photon events, the first equiprobable condition occurs at $T \approx 4 \tau_d$, which corresponds to the center of the pulse, and where the probabilities of detecting both photons in the late bin ($P_{02}$) and one in each bin ($P_{11}$) become equal. This redistribution continues until the second crossing point, when $P_{11} = P_{20}$, is reached at around $T = 12 \tau_d$. At the one-photon level, it can be observed that there exists a value of $T$ where $\text{P}_{10}$ and $\text{P}_{01}$ are also equiprobable, denoted by the dashed vertical line $(T\approx 11.5\tau_d)$. This condition is ideal for preparing quantum states expressed as a superposition of the one-photon components in the Early and Late time bins, $\ket{\psi^{\pm}} = \frac{1}{\sqrt{2}}(\ket{01} \pm \ket{10})$. However, there is still a small two-photon component, arising from the terms $\text{P}_{11}$ and $\text{P}_{02}$, which could act as a source of error when preparing the state $\ket{\psi^{\pm}}$, reducing its fidelity.
\\\\
A more intriguing behavior is observed in the temporal structure for even pulse areas (Fig. \ref{fig:PhotProb2LS}c), where two-photon processes dominate. Just as for odd pulse areas, all photon events within the late time bin remain constant until $T \approx 3 \tau_d$, with the difference that one- and two-photon probabilities are much closer. Furthermore, the temporal structure reveals that the first mixing points for one-photon ($P_{01} = P_{10}$) and two-photon ($P_{02} = P_{20}$) events occur closer in time compared to the odd pulse case. Therefore, the temporal structure of the two-photon emission indicates that it would not be possible to directly prepare NOON states $\frac{1}{\sqrt{2}}(\ket{N0}\pm\ket{0N})$ using a single pulse of even pulse area.
\section{filtered dynamical Resonance Fluorescence}\label{sec:FiltRF}
\begin{figure}[h!]
    \centering
    \includegraphics[width=\columnwidth]{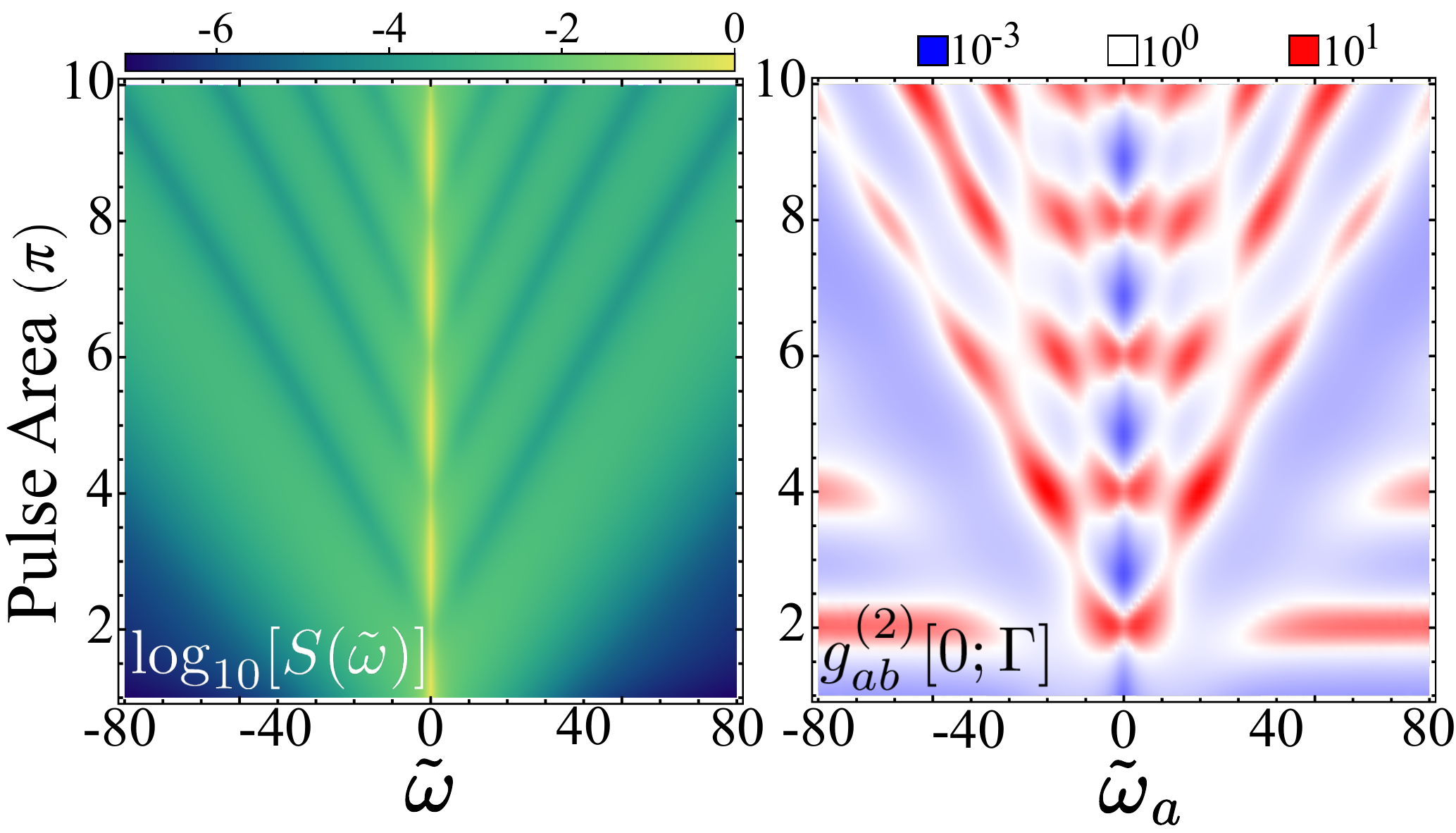}
    \caption{a.) Emission spectrum $S(\Tilde{\omega})$ (Eq.(\ref{Spectrum_eq})) in logarithmic scale of the two-level system as a function of the pulse area ($\Theta / \pi$). b.) Frequency-filtered second-order correlation function $g^{(2)}_{ab}[0 ; \Gamma]$, as a function of the sensor frequency $\Tilde{\omega}_a$ and pulse area. For this figure, the frequency of both sensors is taken to be the same, and sensor bandwidth is set to $\Gamma = 2\gamma_\sigma$.}
    \label{fig:Spectrum2LSPulseArea}
\end{figure}

To measure photons of different frequencies emitted by the dynamically driven two-level system, it is necessary to calculate frequency-resolved correlations. This can be achieved using either the \textit{Sensor Formalism} developed by Del Valle \textit{et al.} \cite{del_valle_theory_2012} or by applying the \textit{Cascaded Formalism} \cite{carmichael_open_1993, gardiner_driving_1993}. We will focus on the first approach, which consists of including $n$ sensors, modeled as external TLS of frequencies $\omega_j$ and decay rates $\Gamma_j$, within the dynamics, allowing them to interact but not alter the system dynamics (i.e. vanishing coupling). This formalism modifies the master equation (\ref{eq:2LSME}) as follows:
\begin{align}    \label{eq:SensorFormalismME} \partial_t\rho &= i\left[\rho,H_\sigma(t)+H_S+ H_{\text{int}}\right]\\\nonumber
&+\frac{\gamma_\sigma}{2}\mathcal{L}_\sigma\rho+\sum_{j}^{n}\frac{\Gamma_j}{2}\mathcal{L}_{\zeta_j}\rho \,,
\end{align}
where $H_\sigma(t)$ is given by Eq.(\ref{eq:dynTLSHam}), $H_S = \sum_{j}^{n}\Tilde{\omega}_j \zeta^{\dagger}_j\zeta_j$ represents the free energy of the sensors and $H_{\text{int}} = \sum_{j}^{n}\epsilon(\sigma^\dagger \zeta_j +\zeta^{\dagger}_j\sigma)$ gives the interaction with the system. Once again, by solving the master equation (\ref{eq:SensorFormalismME}) in a basis that includes the TLS and the sensors, i.e., $\ket{\Psi}=\ket{\alpha}\otimes\prod_{j}^{n}\ket{\eta_j}$ (with $\alpha = G, X$, and $\eta_j = 0, 1$), it is possible to calculate two-time correlation functions of the form $\left\langle A(t)B(t+\tau)C(t)\right\rangle$ for both the TLS ($\sigma$) and sensor operators ($\zeta_j$), to study the emission properties with finite driving at particular frequencies. \\\\
As a first application of this formalism, we compute the emission spectrum of the TLS using the two-time first-order correlation function:
\begin{align}\label{Spectrum_eq}
    S(\Tilde{\omega})=\Re\left\lbrace\int_0^{\infty} \int_0^{\infty} G^{(1)}_\sigma(t, \tau) e^{i \tilde{\omega} \tau} \,d \tau \,d t\right\rbrace,
\end{align}
where $G^{(1)}_{\sigma}(t,\tau)=\left\langle \sigma^\dagger(t+\tau)\sigma(t)\right\rangle$ is the first-order correlation function and $\Tilde{\omega} = \omega - \omega_L$. Equivalently, following the method from Del Valle \textit{et al.} \cite{del_valle_theory_2012}, the emission spectrum can also be computed with the use of the sensor method by using the population of a single sensor in the limit $\Gamma\xrightarrow[]{}0$. The results of the emission spectrum as a function of the pulse area ($\Theta$) are shown in Fig. \ref{fig:Spectrum2LSPulseArea}a. From it, the dominant central frequency line, as well as the characteristic $2n\pi$ side peaks resulting from the Rabi rotations of Dynamically Dressed states \cite{boos_signatures_2024}, are retrieved. 
Now, when considering the statistical properties of the emission, with varying frequency of the emitted photons, the relevant quantity is the time-integrated, cross-correlated second-order correlation function between the sensors, defined as:
\begin{align}
g^{(2)}_{ab}[0 ; \Gamma] = \frac{\iint ( G_{ab}^{(2)}(t,\tau) + G_{ba}^{(2)}(t,\tau) )\,dt\,d\tau}{\left(\int n_a\, dt\right)\left(\int n_b\, dt\right)},\label{g2abzero}
\end{align}
where $n_c(t)=\left\langle \zeta^{\dagger}_c(t) \zeta_c(t)\right\rangle$, for $c\in{a,b}$, represents the mean number of photodetections as measured by sensor $c$. As shown in Fig. \ref{fig:Spectrum2LSPulseArea}b, the emission properties of the two-level system do not exhibit a simple transition between antibunching and bunching for odd and even pulse areas once frequency resolution is incorporated into the detection scheme. The plot reveals that, while oscillatory behavior is linked to the pulse area ($\Theta$), the system displays contrasting photon correlation characteristics in specific frequency regions. For instance, at even values of $\Theta$, broad regions exhibit $g^{(2)}_{ab} < 1$, indicating antibunching. Conversely, bunching behavior is observed at odd pulse areas, particularly in regions away from the central peak emission.\\\\
This feature becomes more evident when calculating $g^{(2)}_{ab}[0 ; \Gamma]$ as a function of the sensor frequencies $\Tilde{\omega}_a$ 
\begin{figure}[h!]
\includegraphics[width=\columnwidth]{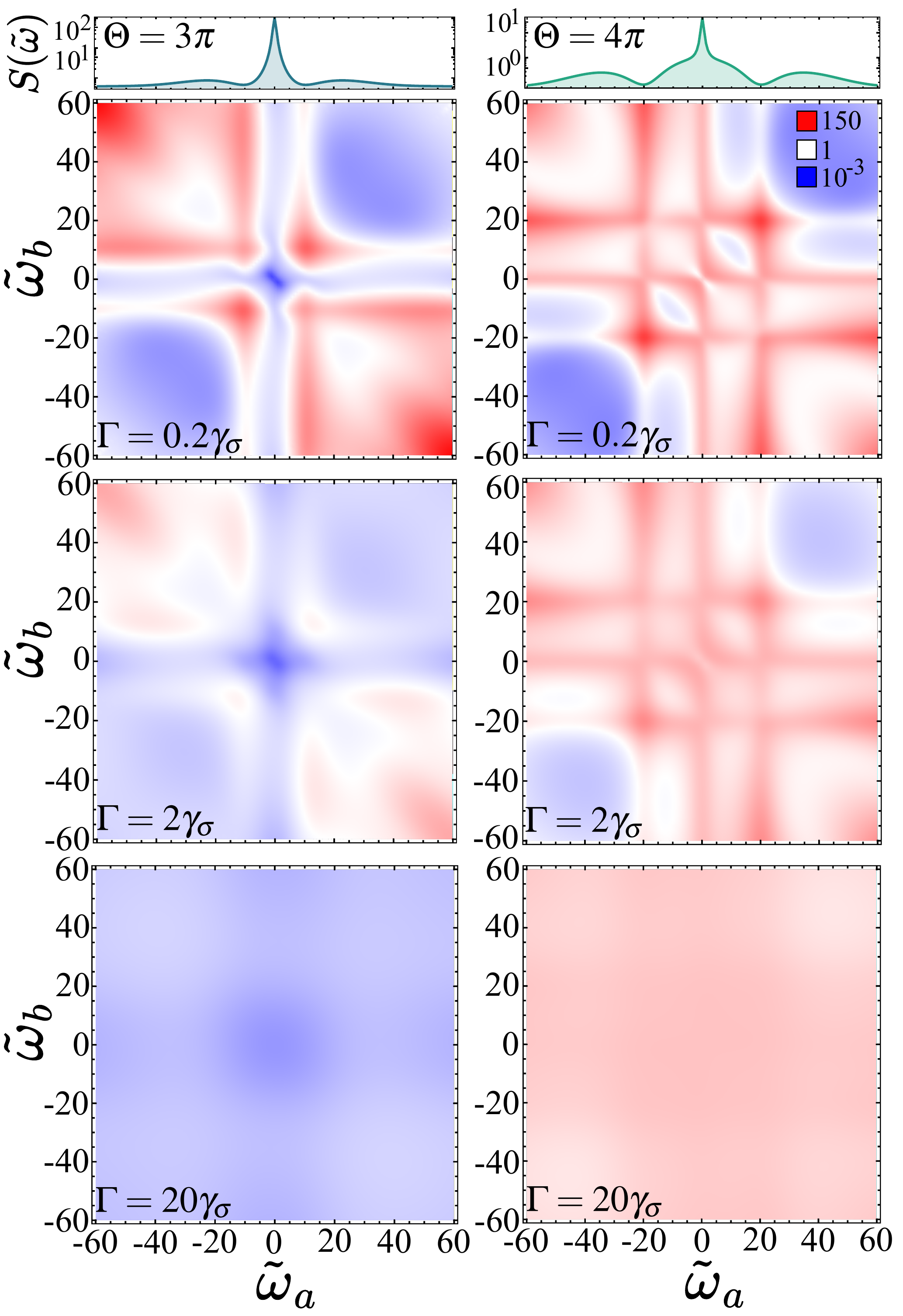}
  \caption{Second order correlation function $g^{(2)}_{ab}[0 ; \Gamma]$ of the detected photons for pulse areas of $3\pi$ (left panel) and $4\pi$ (right panel) as a function of the sensor frequencies for three bandwidths regimes: Sub-linewidth $(\Gamma < \gamma_\sigma)$, Linewidth $(\Gamma = \gamma_\sigma)$ 
 and Broad-linewidth $(\Gamma > \gamma_\sigma)$.}
 \label{fig:g2FreqDependence}
\end{figure}
and $\Tilde{\omega}_b$, as shown in Fig \ref{fig:g2FreqDependence}. In this figure, the values for pulse areas of $3\pi$ and $4\pi$ were calculated for three different regimes of the sensor bandwidths, depending on the ratio of $\Gamma / \gamma_\sigma$ (Appendix \ref{appendix:HigherPulseAreas} shows the effect on higher pulse areas).\\\\
For the case of $3\pi$, the highest degree of antibunching arises from filtering the central peak with sub-linewidth resolution ($\Gamma < \gamma_\sigma$). Moreover, correlating the central peak with any other part of the spectrum preserves a certain degree of antibunching. However, broad regions of bunching are also present. For instance, in the case of photon autocorrelation, where $\Tilde{\omega}_a =\Tilde{\omega}_b$, bunching is observed when filtering the valleys between the central peak and one of the sidepeaks. This remains true when correlating these valleys with photons from different energies, except when the frequency matches the central peak. Notably, these features shared some similarities with the CW case, where this type of strong bunching has its origin in the two-photon transitions between dressed-atom manifolds, called \emph{leapfrog transitions}~\cite{lopez_carreno_photon_2017}. The most significant bunching, however, occurs along the main antidiagonal, where $\Tilde{\omega}_a + \Tilde{\omega}_b = 0$, particularly when filtering opposite tails of the emission. Interestingly, the bunching behavior along this antidiagonal does not increase smoothly from the central peak toward the tails of the emission. Instead, it decays as it approaches the sidepeak maximum, resulting in a small degree of antibunching at that point.
The departure from the continuous-wave behavior suggests that the rate equation model, which successfully describes dressed-state transitions under steady-state driving, may not straightforwardly extend to the pulsed regime, where dynamic spectral features and temporal correlations play a more pronounced role. Furthermore, as the sensor bandwidth increases relative to the decay rate of the two-level system, the strong bunching features gradually diminish and become uncorrelated. Only the valley autocorrelations and cross-correlations between the emission tails remain noticeable. Once the sensor bandwidth, $\Gamma$, exceeds the system's decay rate, $\gamma_\sigma$, the expected antibunching behavior of emission under $3\pi$ driving is restored.\\\\
Now, as identified by Fischer \textit{et al.} \cite{fischer_signatures_2017}, the use of a driving pulse with a duration shorter than the emission lifetime, with an even pulse area, leads to the possibility of \textit{re-excitation}, i.e., during the system–pulse interaction, the pulse may induce an initial emission early in the interaction, and with the remainder of the pulse, it can re-excite the system, generating the emission of additional photons. This leads to a dominant amount of two-photon emission, resulting in bunched photon statistics.
 As illustrated in the right panel of Fig. \ref{fig:g2FreqDependence}, the most prominent transitions largely exhibit this behavior, where correlations between the central peak and the rest of the spectrum remain bunched. This also holds for correlations between photons emitted from the valley between the central peak and the sidepeaks and the rest of the spectrum. Interestingly, the autocorrelations from these valleys exhibit the highest degree of bunching, surpassing even that of the central peak. However, there are some notable regions where antibunching persists at this pulse area, with the most prominent example being the autocorrelations from the sidepeaks. Another notable feature is observed at sub-linewidth resolution when filtering the shoulders of the broad central spectrum, attributed to the emission of a supernatural linewidth photon within the pulse excitation. However, the degree of antibunching, in this case, is smaller than when filtering the sidepeaks and becomes unresolvable as the bandwidth increases, giving in the end the complete bunched nature of the two-photon emission.\\\\
We complete our analysis of spectral correlations in pulsed resonance fluorescence by examining the effect of driving the two-level system with an off-resonant laser. This configuration enables spectral separation between the laser frequency and the two-level system (TLS) transition, allowing us to explore how detuning alters the photon correlation landscape. Specifically, we consider a detuning of
$\tilde{\omega}_\sigma = 20\gamma_\sigma$, as shown in Fig.(\ref{fig:g2FreqDependence_Detuning}), where laser position is indicated by the dashed lines.\\\\ The most prominent effect of detuning on the emission statistics is that antibunching, for both odd and even pulse areas, now resides over the laser's central frequency rather than on the peak corresponding to the TLS transition. Additionally, the intensity of the TLS transition peak is significantly diminished. Furthermore, several antibunching regions emerge for both pulse areas, corresponding to correlations between the same resolvable peaks or cross-correlations between the TLS transition frequency and the emergent dynamical sidepeaks.
\begin{figure}[h!]
\includegraphics[width=\columnwidth]{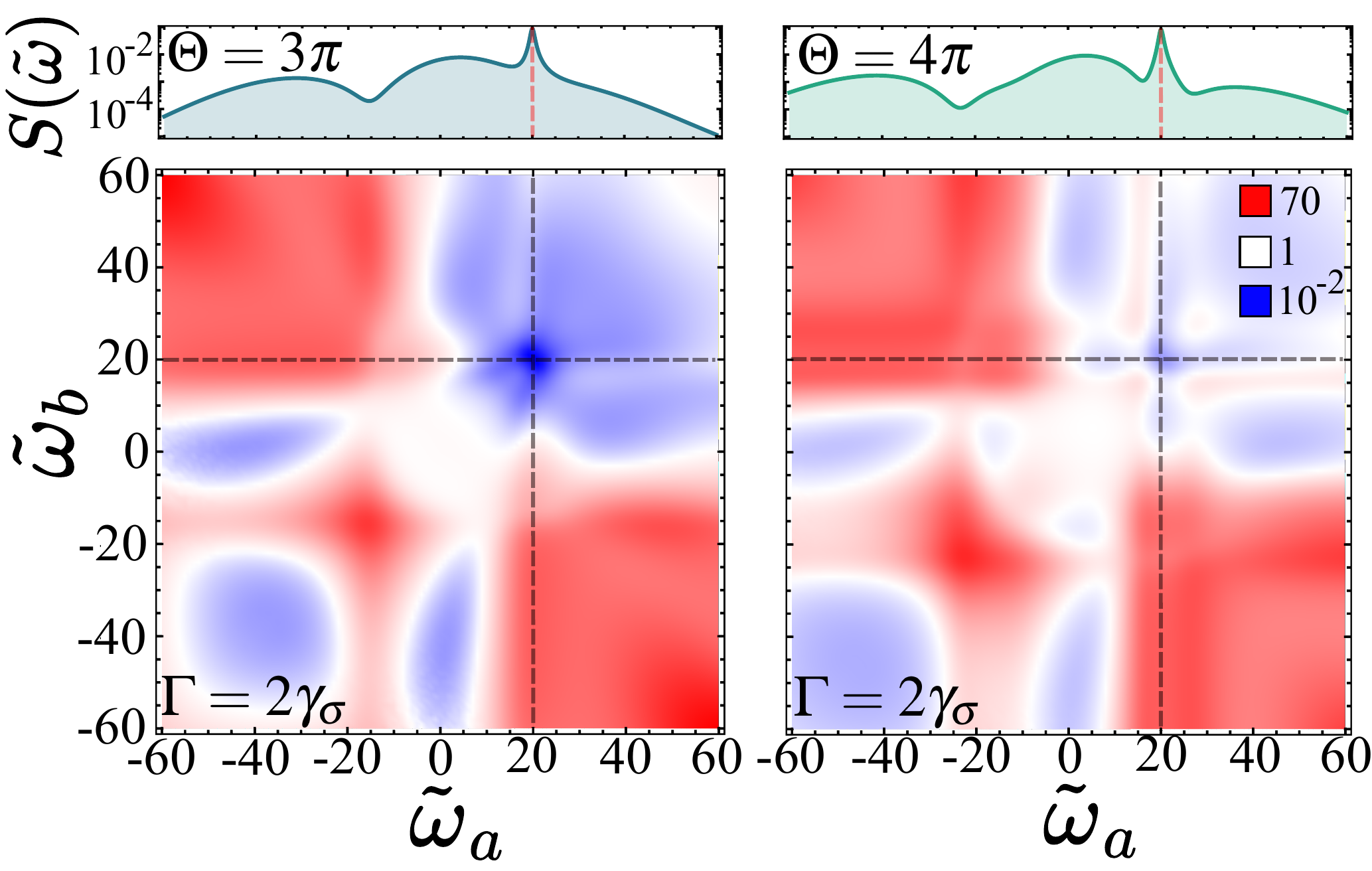}
  \caption{Second order correlation function $g^{(2)}_{ab}[0 ; \Gamma]$ of the detected photons for pulse areas of $3\pi$ (left panel) and $4\pi$ (right panel) as a function of the sensor frequencies for off-resonant driving, with $\Tilde{\omega}_\sigma = 20\gamma_\sigma$,  highlighted on the spectrum and correlation maps as dashed lines.}
\label{fig:g2FreqDependence_Detuning}
\end{figure}
\\When the detuning is moderate, such that the spectral peaks are not fully separated, overlapping regions lead to a mixing of photon statistics from adjacent peaks. For example, at a pulse area of $3\pi$, there is no clearly defined valley between the laser and TLS peaks, which contrasts with the resonant case where bunching is prominent in the valley region. As a result, bunching is diminished under detuning. In contrast, for a pulse area of $4\pi$, three distinct valleys become resolvable (compared to two in the resonant case), giving rise to additional bunching lines in the frequency-resolved correlation maps.
\begin{figure*}
  \includegraphics[width=\textwidth]{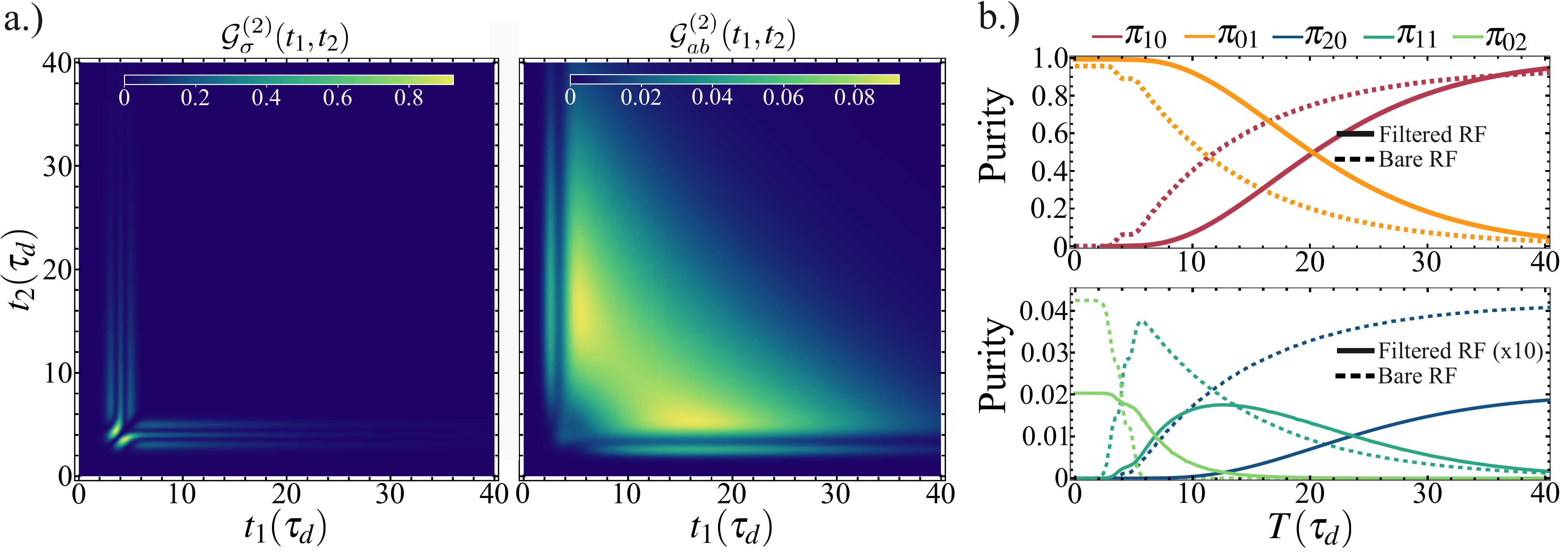}
  \caption{a.) Two timed second-order correlation functions in the case of bare (left) and filtered central peak (right) emission of a pulse area $\Theta = 3\pi$. \textit{Parameters:} $\tilde{\omega}_{a,b} = 0$, $t_0 = 4\tau_d$, $\Gamma = 2\gamma_\sigma$ b.) Time bin purities (\ref{eq:timebin}) for the bare and filtered emission of a two-level system, related to the second-order correlation plots given in a. Two-photon filtered purities are magnified by a factor of 10 for comparison.}
\end{figure*}\label{fig:Timebinpurity3Pi}
\\\\As a final part of our work, we show how the photon time bins probability distributions change for the case of frequency filtering. This is analogous to what was found in Fig.(\ref{fig:PhotProb2LS}b) but with specifically selected frequencies of the TLS dynamical spectrum. By inspection of Fig.(\ref{fig:g2FreqDependence}), we can choose particular frequencies where we ensure that we have the desired one photon (antibunched) and two photons (bunched) statistics. Moreover, we investigate if by adequately selecting these frequencies, we can alter the one and two-photon probabilities shown for bare resonance fluorescence.\\\\
For that matter, we applied the two-mode probability equations given in Appendix \ref{appendix:PhotCount}, and used them to compute the \textit{time bin purities} ($\pi_{mn}$), which are defined as the time bin probabilities employed on Figures \ref{fig:PhotProb2LS}b,c, but renormalizing them by removing the vacuum contribution, 
\begin{align}\label{eq:timebin}
    \pi_{mn} = \text{P}_{mn}/ \sum_{m+n>0} \text{P}_{mn}.
\end{align}
The resulting behavior of the time bin purities for a pulse area of $3\pi$ is shown in Fig.~(5b), for bare and filtered emission of resonance fluorescence, where for the latter the filter was placed over the central peak ($\tilde{\omega}_{a,b} = 0$) with a bandwidth $\Gamma = 2\gamma_\sigma$. In the first part of this graph (Fig. 5a), we highlight the differences of the second-order correlation function for the bare $(\mathcal{G}^{(2)}_\sigma)$ and filtered $(\mathcal{G}^{(2)}_{ab})$ schemes. For the bare case, where all photons are collected regardless of their frequency, the overall intensity is higher because of contributions from the entire emission spectrum, which results in higher intensity correlations. Moreover, there is a clear temporal structure in the correlation map, which indicates that the emission of the photons is done into different time windows. This is due to the nature of the driving on the TLS, where the emission of the first photon does not constrain that of later photons, which are emitted randomly over the natural timescale of the TLS relaxation. In contrast, spectral filtering isolates photons near the central frequency, reducing the total photon flux. Nevertheless, despite this reduction and that the temporal structure is now delocalized due to the time uncertainty introduced by the filters, the correlation map   exhibits the enhancement of simultaneous detections, as seen in the pronounced diagonal structure of $\mathcal{G}^{(2)}_{ab}(t_1,t_2)$. 
This feature reflects the increased likelihood that photon pairs are emitted at equal times, as a consequence of the enhanced temporal coherence introduced by spectral filtering. This property has also been recently recognized as a signature of stimulated emission of a TLS driven with non-classical light \cite{hansen_non-classical_2024}, and has been identified as a feature of spontaneous two-photon emission \cite{liu_quantum_2024}.\\\\
Fig.~(5b) presents the time-bin purities associated with the bare and filtered photon correlations shown in part (a). The reduced photon flux significantly influences the probability of two-photon events, as reflected in the two-photon purities now being on the order of $10^{-3}$. However, since the filter is applied to the central peak, where the spectral intensity is at its maximum, using a narrow filter of width $2\gamma_\sigma$ results in an estimated intensity reduction by approximately $70\%$. While this reduction affects the overall photon flux, it can be compensated for by extending the integration time. 
Additionally, this approach offers the benefit of potentially higher fidelity in time-bin entangled state preparation \cite{bracht_theory_2024}, as it helps suppress higher photon probabilites. 
\begin{figure}
    \centering
    \includegraphics[width = \columnwidth]{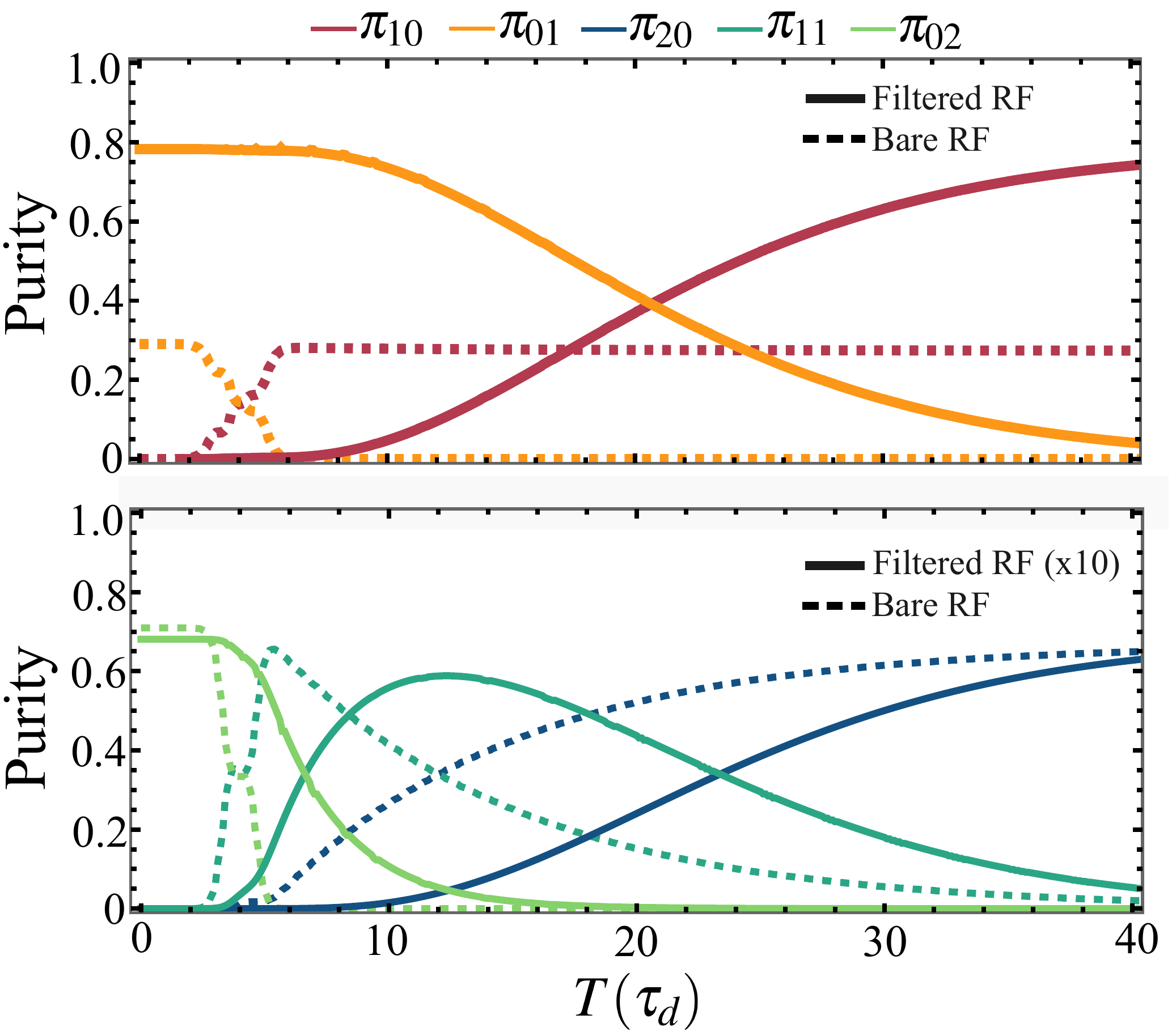}
    \caption{Time bin purities for the bare and filtered emission of a two-level system for a pulse area of $\Theta = 4\pi$. Two-photon filtered purities are magnified by a factor of 10 for comparison. \textit{Parameters:} $\tilde{\omega}_{a,b} = 0$, $\Gamma = 2\gamma_\sigma$.}
    \label{Timebinpurity4Pi}
\end{figure}
\\\\Finally, Fig.(6) illustrates the impact of spectral filtering on the time-bin purities for an even pulse area of $4\pi$. The upper panel depicts the purities $\pi_{10}$ and $\pi_{01}$ (associated with single-photon events), while the lower panel highlights the purities $\pi_{20}$, $\pi_{11}$, and $\pi_{02}$ (associated with two-photon events). The effects of applying a narrow spectral filter with $\Gamma = 2\gamma_\sigma$ are compared to the bare resonance fluorescence (dashed lines). 
As shown, spectral filtering effectively reverses the characteristic tendency of even pulse areas, where $\pi_2 > \pi_1$. By isolating the central spectral peak, the single-photon purities $\pi_{10}$ and $\pi_{01}$ are significantly enhanced, suppressing the contributions from two-photon processes. This improvement is particularly evident at intermediate values of the time-bin splitting parameter $T$, e.g. at the value where the filtered single photon time bin purities are equiprobable ($T\approx 20\tau_d$), where the filtering leads to a reduction in $\pi_{20}$ and $\pi_{02}$ by more than an order of magnitude. Furthermore, the lower panel reveals that for the filtered case, the two-photon purities $\pi_{20}$, $\pi_{11}$, and $\pi_{02}$ become nearly negligible, reflecting the diminished likelihood of multi-photon emissions within the filtered spectral range. This is consistent with the suppression of sidebands outside the filter's bandwidth, as shown in the spectral analysis.

\section{Conclusion}
In this work, we have advanced the understanding of photon statistics in pulsed quantum light sources by systematically exploring spectral correlations in dynamically driven two-level systems. We have thereby extended the formalism to compute frequency-resolved
photon correlations, mainly utilized in the continuous-wave regime, to time-dependent excitation. By calculating time-integrated, frequency-resolved photon correlations, we demonstrated that spectral filtering is a fundamental tool to tailor photon emission properties,
which are essentially linked to the behavior of the photon-counting probabilities. \\\\
Our methodology, which combines the sensor formalism with a quantum regression theorem-based approach, provides a versatile and accessible platform for analyzing frequency correlations in pulsed systems. Likewise, this formalism provides an efficient way of obtaining time-integrated
correlations, whose brute-force computation traditionally relies on iteratively applying the Quantum Regression Theorem to the required order ($N-1$ times for any $N$-time correlation function) and subsequently integrating over all times. Unlike prior studies focused on continuous-wave excitation, our results highlight the intricate interplay between dynamic spectral features and temporal coherence under pulsed driving. Notably, spectral filtering breaks the direct dependence of photon statistics on pulse area, allowing transitions between antibunching and bunching regimes based on the selection of specific frequency and time windows. This tunability opens avenues for designing quantum light sources with on-demand photon statistics, and paves the way for new protocols based on time-bin and frequency
entanglement.\\\\
More specifically, our findings reveal that isolating the central spectral peak of resonance fluorescence not only enhances single-photon purity but also suppresses multi-photon noise, even under conditions traditionally favoring multi-photon emission (e.g., at even pulse areas). This provides a pathway for controlling time-bin-encoded quantum states, which is critical for applications in quantum communication and computing.
\\\\
Overall, in this work we illustrate with the simplest, yet quintessential, example---the two-level system---the capabilities of the methodology presented here. Beyond the seek of a better single-photon emission, the underlying correlations seem to indicate that there is much room left to explore and, as for the CW case, multiphoton emission might be possible by exploiting such correlations, such as the emission of $N$-photon bundles~\cite{sanchezmunoz2014a,zou2023}.
The type of quantum-optical systems we can study is, of course, not limited to
Resonance Fluorescence and the method can be applied, in principle, to any system, provided that the Quantum Regression Theorem holds. In the same way, the formalism can be extended to include more degrees of freedom, such as polarization, corresponding to different transitions or de-excitation pathways; more than two time bins, especially useful to study cluster state generation; more detectors, each of them with different frequencies and filter widths; and more complex excitation schemes, like using a train of pulses, or feeding the system with quantum light \cite{carreno_cascaded_2024, kiilerich_2020}.

\section{Acknowledgements}
S. Bermúdez-Feijóo acknowledges J.N Claro and P. Kallert for fruitful discussions. This work is supported by the ERC grant (LiNQs, 101042672), the Deutsche Forschungsgemeinschaft (German Research Foundation) through the transregional collaborative research center TRR142/3-2022 (231447078)
and the project CNLG (MU 4215/4-1), Germany’s Excellence Strategy – EXC-2111 – 390814868 (MCQST) and the German Federal Ministry of Education and Research (BMBF) through the project “QR.N” (16KIS2206).

\bibliographystyle{unsrt}
\bibliography{References}

\begin{thebibliography}{10}

\bibitem{mollow_power_1969}
B.~R. Mollow.
\newblock Power {Spectrum} of {Light} {Scattered} by {Two}-{Level} {Systems}.
\newblock {\em Phys. Rev.}, 188(5):1969--1975, December 1969.

\bibitem{flagg_resonantly_2009}
E.~B. Flagg, A.~Muller, J.~W. Robertson, S.~Founta, D.~G. Deppe, M.~Xiao,
  W.~Ma, G.~J. Salamo, and C.~K. Shih.
\newblock Resonantly driven coherent oscillations in a solid-state quantum
  emitter.
\newblock {\em Nature Phys}, 5(3):203--207, March 2009.

\bibitem{lopez_carreno_photon_2017}
J.C. López~Carreño, E.~Del~Valle, and F.~P. Laussy.
\newblock Photon {Correlations} from the {Mollow} {Triplet}.
\newblock {\em Laser \& Photonics Reviews}, 11(5):1700090, September 2017.

\bibitem{peiris_two-color_2015}
M.~Peiris, B.~Petrak, K.~Konthasinghe, Y.~Yu, Z.~C. Niu, and A.~Muller.
\newblock Two-color photon correlations of the light scattered by a quantum
  dot.
\newblock {\em Phys. Rev. B}, 91(19):195125, May 2015.

\bibitem{gonzalez-tudela_two-photon_2013}
A.~Gonzalez-Tudela, F.~P. Laussy, C.~Tejedor, M.~J. Hartmann, and E.~del Valle.
\newblock Two-photon spectra of quantum emitters.
\newblock {\em New J. Phys.}, 15(3):033036, March 2013.

\bibitem{nieves_third-order_2020}
Y.~Nieves and A.~Muller.
\newblock Third-order photon cross-correlations in resonance fluorescence.
\newblock {\em Phys. Rev. B}, 102(15):155418, October 2020.

\bibitem{del_valle_theory_2012}
E.~Del~Valle, A.~Gonzalez-Tudela, F.~P. Laussy, C.~Tejedor, and M.~J. Hartmann.
\newblock Theory of {Frequency}-{Filtered} and {Time}-{Resolved} {N} -{Photon}
  {Correlations}.
\newblock {\em Phys. Rev. Lett.}, 109(18):183601, October 2012.

\bibitem{kim_unlocking_2024}
K.~Sang~Kyu, E.~Zubizarreta~Casalengua, K.~Boos, F.~Sbresny, C.~Calcagno,
  H.~Riedl, J.~J. Finley, C.~Antón-Solanas, F.~P. Laussy, K.~Müller,
  L.~Hanschke, and E.~Del~Valle.
\newblock Unlocking multiphoton emission from a single-photon source through
  mean-field engineering, November 2024.
\newblock arXiv:2411.10441 [quant-ph].

\bibitem{lopez_carreno_entanglement_2024}
J.~C. López~Carreño, S.~Bermúdez~Feijoo, and M.~Stobińska.
\newblock Entanglement in {Resonance} {Fluorescence}.
\newblock {\em npj Nanophoton.}, 1(1):3, April 2024.

\bibitem{moelbjerg_resonance_2012}
A.~Moelbjerg, P.~Kaer, M.~Lorke, and J.~Mørk.
\newblock Resonance {Fluorescence} from {Semiconductor} {Quantum} {Dots}:
  {Beyond} the {Mollow} {Triplet}.
\newblock {\em Phys. Rev. Lett.}, 108(1):017401, January 2012.

\bibitem{boos_signatures_2024}
K.~Boos, S.~K. Kim, T.~Bracht, F.~Sbresny, J.~M. Kaspari, M.~Cygorek, H.~Riedl,
  F.~W. Bopp, W.~Rauhaus, C.~Calcagno, J.~J. Finley, D.~E. Reiter, and
  K.~Müller.
\newblock Signatures of {Dynamically} {Dressed} {States}.
\newblock {\em Phys. Rev. Lett.}, 132(5):053602, January 2024.

\bibitem{liu_dynamic_2024}
S.~Liu, C.~Gustin, H.~Liu, X.~Li, Y.~Yu, H.~Ni, Z.~Niu, S.~Hughes, X.~Wang, and
  J.~Liu.
\newblock Dynamic resonance fluorescence in solid-state cavity quantum
  electrodynamics.
\newblock {\em Nat. Photon.}, 18(4):318--324, April 2024.

\bibitem{senellart_high-performance_2017}
P.~Senellart, G.~Solomon, and A.~White.
\newblock High-performance semiconductor quantum-dot single-photon sources.
\newblock {\em Nature Nanotech}, 12(11):1026--1039, November 2017.

\bibitem{santori_triggered_2001}
C.~Santori, M.~Pelton, G.~Solomon, Y.~Dale, and Y.~Yamamoto.
\newblock Triggered {Single} {Photons} from a {Quantum} {Dot}.
\newblock {\em Phys. Rev. Lett.}, 86(8):1502--1505, February 2001.

\bibitem{santori_indistinguishable_2002}
C.~Santori, D.~Fattal, J.~Vučković, G.~S. Solomon, and Y.~Yamamoto.
\newblock Indistinguishable photons from a single-photon device.
\newblock {\em Nature}, 419(6907):594--597, October 2002.

\bibitem{he_-demand_2013}
Y.-M. He, Y.~He, Y.-J. Wei, D.~Wu, M.~Atatüre, C.~Schneider, S.~Höfling,
  M.~Kamp, C.-Y. Lu, and J.-W. Pan.
\newblock On-demand semiconductor single-photon source with near-unity
  indistinguishability.
\newblock {\em Nature Nanotech}, 8(3):213--217, March 2013.

\bibitem{loredo_generation_2019}
J.~C. Loredo, C.~Antón, B.~Reznychenko, P.~Hilaire, A.~Harouri, C.~Millet,
  H.~Ollivier, N.~Somaschi, L.~De~Santis, A.~Lemaître, I.~Sagnes, L.~Lanco,
  A.~Auffèves, O.~Krebs, and P.~Senellart.
\newblock Generation of non-classical light in a photon-number superposition.
\newblock {\em Nat. Photonics}, 13(11):803--808, November 2019.

\bibitem{vajner_towards_2024}
D.~A. Vajner, N.~D. Kewitz, M.~von Helversen, S.~C. Wein, Y.~Karli, F.~Kappe,
  V.~Remesh, S.~F. Covre~da Silva, A.~Rastelli, G.~Weihs, C.~Anton-Solanas, and
  T.~Heindel.
\newblock Towards {Photon}-{Number}-{Encoded} {High}-dimensional {Entanglement}
  from a {Sequentially} {Excited} {Quantum} {Three}-{Level} {System}, 2024.
\newblock Version Number: 1.

\bibitem{fischer_signatures_2017}
K.~A. Fischer, L.~Hanschke, J.~Wierzbowski, T.~Simmet, C.~Dory, J.~J. Finley,
  J.~Vučković, and K.~Müller.
\newblock Signatures of two-photon pulses from a quantum two-level system.
\newblock {\em Nature Phys}, 13(7):649--654, July 2017.

\bibitem{konthasinghe_correlations_2015}
K.~Konthasinghe, M.~Peiris, B.~Petrak, Y.~Yu, Z.~C. Niu, and A.~Muller.
\newblock Correlations in pulsed resonance fluorescence.
\newblock {\em Opt. Lett.}, 40(8):1846, April 2015.

\bibitem{carreno_cascaded_2024}
J.~C. Carreño~López.
\newblock Cascaded {Single} {Photons} from {Pulsed} {Quantum} {Excitation},
  November 2024.
\newblock arXiv:2411.16539 [quant-ph].

\bibitem{cardoso_impact_2025}
F.~Redivo~Cardoso, J.~Lee, R.~Checchinato, J.-H. Littmann, M.~De~Gregorio,
  S.~Höfling, C.~Schneider, C.~J. Villas-Boas, and A.~Predojević.
\newblock Impact of temporal correlations, coherence, and postselection on
  two-photon interference.
\newblock {\em Phys. Rev. Research}, 7(1):013190, February 2025.

\bibitem{kimble_quantum_2008}
H.~J. Kimble.
\newblock The quantum internet.
\newblock {\em Nature}, 453(7198):1023--1030, June 2008.

\bibitem{carmichael_open_1993}
Howard Carmichael.
\newblock {\em An {Open} {Systems} {Approach} to {Quantum} {Optics}: {Lectures}
  {Presented} at the {Université} {Libre} de {Bruxelles} {October} 28 to
  {November} 4, 1991}, volume~18 of {\em Lecture {Notes} in {Physics}
  {Monographs}}.
\newblock Springer Berlin Heidelberg, Berlin, Heidelberg, 1993.

\bibitem{breuer_theory_2007}
H.-P. Breuer and F.~Petruccione.
\newblock {\em The {Theory} of {Open} {Quantum} {Systems}}.
\newblock Oxford University PressOxford, 1 edition, January 2007.

\bibitem{gardiner_quantum_2000}
C.~W. Gardiner and P.~Zoller.
\newblock {\em Quantum noise: a handbook of {Markovian} and non-{Markovian}
  quantum stochastic methods with applications to quantum optics}.
\newblock Number~56 in Springer series in synergetics. Springer, Berlin New
  York, 2nd enl. ed edition, 2000.

\bibitem{molmer_monte_1993}
K.~Mølmer, Y.~Castin, and J.~Dalibard.
\newblock Monte {Carlo} wave-function method in quantum optics.
\newblock {\em J. Opt. Soc. Am. B}, 10(3):524, March 1993.

\bibitem{lopez_carreno_frequency-resolved_2018}
J.~C. López~Carreño, E.~del Valle, and F.~P. Laussy.
\newblock Frequency-resolved {Monte} {Carlo}.
\newblock {\em Sci Rep}, 8(1):6975, May 2018.

\bibitem{gardiner_driving_1993}
C.~W. Gardiner.
\newblock Driving a quantum system with the output field from another driven
  quantum system.
\newblock {\em Phys. Rev. Lett.}, 70(15):2269--2272, April 1993.

\bibitem{hansen_non-classical_2024}
L.~M. Hansen, F.~Giorgino, L.~Jehle, L.~Carosini, J.~C. López~Carreño,
  I.~Arrazola, P.~Walther, and J.~C. Loredo.
\newblock Non-classical excitation of a solid-state quantum emitter, 2024.
\newblock Version Number: 1.

\bibitem{liu_quantum_2024}
J.~Liu, S.~Liu, Y.~Wang, Y.~Saleem, X.~Li, H.~Liu, C.-A. Yang, J.~Yang, H.-Q.
  Ni, Z.~Niu, Y.~Meng, X.~Hu, Y.~Yu, X.-H. Wang, and M.~Cygorek.
\newblock Quantum {Nature} of {Spontaneous} {Two}-{Photon} {Emission} in
  {Semiconductor} {Cavity} {Quantum} {Electrodynamics}, December 2024.

\bibitem{bracht_theory_2024}
T.~K. Bracht, F.~Kappe, M.~Cygorek, T.~Seidelmann, Y.~Karli, V.~Remesh,
  G.~Weihs, V.~M. Axt, and D.~E. Reiter.
\newblock Theory of time-bin-entangled photons from quantum emitters.
\newblock {\em Phys. Rev. A}, 110(6):063709, December 2024.

\bibitem{sanchezmunoz2014a}
C.~S{\'a}nchez Mu{\~n}oz, E.~Del~Valle, A.~Gonz{\'a}lez Tudela, K.~M{\"u}ller,
  S.~Lichtmannecker, M.~Kaniber, C.~Tejedor, J.~J. Finley, and F.~P. Laussy.
\newblock Emitters of {{N-photon}} bundles.
\newblock {\em Nature Photonics}, 8(7):550--555, July 2014.

\bibitem{zou2023}
F~Zou, Y~Li, and J-Q Liao.
\newblock Dynamical {{N-photon}} bundle emission.
\newblock {\em New Journal of Physics}, 25(4):043027, April 2023.

\bibitem{kiilerich_2020}
A.~H. Kiilerich and K.~M{\o}lmer.
\newblock Quantum interactions with pulses of radiation.
\newblock {\em Phys. Rev. A}, 102(2), August 2020.

\end{thebibliography}

\clearpage

\begin{widetext}

\appendix

\section{Quantum Regression Theorem}
To solve the correlations shown before, we need to make use of the Quantum Regression theorem. For such purpose, we define an
observable vector $\vec{c}$ whose averages, namely $\av{\vec{c}}$,
follow the equation of motion, 
\begin{equation}
    \label{eq:ct-EqM}
    \partial_t \av{\vec{c} (t)} = M(t) \av{\vec{c} (t)} \,,
\end{equation}
which can be derived straight from the Master Equation, namely $\dot{\rho} = \mathcal{L} \rho$ (where $\mathcal{L}$
represents the Liouvillian superoperator),
by multiplying $\vec{c}$ and then taking the trace, i.e.,
$\Tr{\dot{\rho} \vec{c}} = \Tr{(\mathcal{L}\rho) \vec{c}}$.
Then, we define the two-time correlation vector 
$\vec{v} (t,t') = \av{A(t) \vec{c}(t') B(t)}$. Provided that
$t'>t$ and the elements $A \vec{c} B$ are all normally ordered
operators, then $\vec{v} (t,t')$ fulfills
\begin{equation}
    \partial_{t'} \vec{v} (t,t') = M(t') \vec{v} (t,t') \,.
\end{equation}
This is the essence of the QRT.
Alternatively, if we parameterize $t'=t + \tau$, the previous equation
reads
\begin{equation}
    \partial_{\tau} \vec{v} (t,\tau) = M(t+\tau) \vec{v} (t,\tau) \,.
\end{equation}
In particular, we will consider the correlation vectors of the type
$\vec{v}_{a_i} (t,t') = \av{a_i^{\dagger} (t) \vec{c}(t') a_i(t)}$
and, additionally, we will make use of the matrix $\mathcal{C}_{a_i}$
that maps the observable vector $\vec{c}$ into $a_i^{\dagger} \vec{c} \, a_i$.\\\\ 
An additional note is that the operator vector $\vec{c}$ may contain either a finite or an infinite number of elements. For instance, in the case of resonance fluorescence, the dynamics are fully captured by the finite vector $\vec{c} = (1, \sigma, \ud{\sigma}, \ud{\sigma} \sigma)^\mathrm{T}$. However, when an infinite set of operators is required, truncation must be applied. The number of operators needed to ensure physically meaningful and fully converged solutions depends on the specific parameters of the system. Generally, we include the identity operator as the 0-th element of the operator vector $\vec{c}$.

\section{Derivation of the equations of motion}

We start by rewriting the integrals $\int_0^T dt_1 \int_0^T dt_2$ as
$\int_0^\infty dt_1 \int_0^\infty dt_2 \, \theta(T-t_1)\theta(T-t_2)$.
Then, the integrated correlation functions in Eq.~\eqref{eq:IntegratedG2_def2} read now

\begin{equation}
    G_{a_1 \rightarrow a_2}^{(2)}[0,T] =
    \int_0^\infty \int_0^\infty \, \theta(T-t_1)\theta(T-t_2)
    \theta(t_2-t_1)\mathcal{G}^{(2)}_{a_1 \rightarrow a_2} (t_1,t_2) 
    \, dt_1 dt_2 \,,
\end{equation}
Now we derive with respect of $T$.
We use Leibniz integral rule to differentiate under the integral sign
and, afterwards, the Leibniz product and the fact that $\partial_T \theta(T-t_i) 
= \delta(T - t_i)$ (for $i = 1, 2$), to eventually obtain
\begin{align}
   \partial_T \, G_{a_1 \rightarrow a_2}^{(2)}[0,T] = & \
    \int_0^\infty \int_0^\infty \, \big[\delta(T-t_1)\theta(T-t_2) +
    \theta(T-t_1)\delta(T-t_2) \big]
    \theta(t_2-t_1)\mathcal{G}^{(2)}_{a_1 \rightarrow a_2} (t_1,t_2) 
    \, dt_1 dt_2 \nonumber \\
    = & \ \int_0^\infty \theta(T-t_2) \theta(t_2-T) \mathcal{G}^{(2)}_{a_1 \rightarrow a_2} (T,t_2) 
    \, dt_2 + 
    \int_0^\infty \theta(T-t_1) \theta(T - t_1) \mathcal{G}^{(2)}_{a_1 \rightarrow a_2} (t_1,T) dt_1
    \,.
\end{align}
The product of the Heaviside functions in the first integral is zero, as
the inequalities $T - t_2 > 0$ and $t_2 - T > 0$ are never fulfilled simultaneously,
whereas the second product is trivially simplified as $\theta(T-t_1)$.
Therefore, it simply yields
\begin{equation}
\partial_T \, G_{a_1 \rightarrow a_2}^{(2)}[0,T] = 
\int_0^\infty \theta(T-t_1) \, \mathcal{G}^{(2)}_{a_1 \rightarrow a_2} (t_1,T) \, dt_1 \,.
\end{equation}
To solve this integral, we need to define the quantity
\begin{equation}
\label{eq:Va1def}
    \vec{V}_{a_1} (T) = \int_0^\infty \theta(T - t_1) \vec{v}_{a_1} (t_1,T) \, dt_1 \,,
\end{equation}
where we remind that $\vec{v}_{a_1} (t,t') = \av{a_i^{\dagger} (t) \vec{c}(t') a_i(t)}$.
It is easy to see that $[\vec{V}_{a_1} (T)]_{i_2} = G_{a_1 \rightarrow a_2}^{(2)}[0,T]$, 
provided that the $i_2$-th element of $\vec{c}$ is $a^\dagger_2 a_2$ (where $[]_i$ denotes the $i$-th
element of a vector).
We repeat the same procedure and derive again with respect of $T$
\begin{equation}
   \partial_T \vec{V}_{a_1} (T) = \int_0^\infty \big[ \delta(T - t_1) \vec{v}_{a_1} (t_1,T) 
   + \theta(T - t_1) \partial_T \vec{v}_{a_1} (t_1,T) \big]
   \, dt_1 \,,
\end{equation}
which, after making use of the QRT, leads to
\begin{equation}
   \partial_T \vec{V}_{a_1} (T) =  \int_0^\infty  \theta(T - t_1) M(T) \vec{v}_{a_1} (t_1,T)
   \, dt_1 +
   \vec{v}_{a_1} (T,T) \,.
\end{equation}
The matrix $M(T)$ does not depend on $t_1$ and can be taken outside the integral
and, introducing the matrix $\mathcal{C}_{a_1}$, we write 
$\vec{v}_{a_1} (T,T) = \av{(a_1^\dagger \vec{c} \, a_1)(T)} = \mathcal{C}_{a_1} \av{\vec{c}(T)}$.
Ultimately, we get to the final expression
\begin{equation}
   \partial_T \vec{V}_{a_1} (T) = 
   M(T) \vec{V}_{a_1} (T) + \mathcal{C}_{a_1} \av{\vec{c}(T)}
   \,.
\end{equation}
%
%
%
Summarizing, the equations of motion are
\begin{subequations}
\begin{align}
  & \partial_T \, G_{a_1 \rightarrow a_2}^{(2)}[0,T] = [\vec{V}_{a_1} (T)]_{i_2} \,, \\
  & \partial_T \vec{V}_{a_1} (T) = M(T) \vec{V}_{a_1} (T) + \mathcal{C}_{a_1} \av{\vec{c}(T)} \,, \\
  & \partial_T \av{\vec{c}(T)} = M(T) \av{\vec{c}(T)} \,,
\end{align}
\end{subequations}
with the initial conditions $G_{a_1 \rightarrow a_2}^{(2)}[0,0] = \vec{V}_{a_1} (0) = 0$
and $\av{\vec{c} (0)} = \Tr{\rho(0) \vec{c}}$ determined by the initial quantum state $\rho (0)$.
Equivalently, for $G_{a_1 \rightarrow a_2}^{(2)}[0,T]$ the equations are
\begin{subequations}
\begin{align}
  & \partial_T \, G_{a_2 \rightarrow a_1}^{(2)}[0,T] = [\vec{V}_{a_2} (T)]_{i_1} \,, \\
  & \partial_T \vec{V}_{a_2} (T) = M(T) \vec{V}_{a_2} (T) + \mathcal{C}_{a_2} \av{\vec{c}(T)} \,, \\
  & \partial_T \av{\vec{c}(T)} = M(T) \av{\vec{c}(T)} \,,
\end{align}
\end{subequations}
with $i_1$ such that $[\vec{c}]_{i_1} = a_1^\dagger a_1$ and initial conditions 
$G_{a_2 \rightarrow a_1}^{(2)}[0,0] = \vec{V}_{a_2} (0) = 0$.

Finally, for the symmetric integrated correlation function, following from Eq.~\eqref{eq:intG2_12}, we have
\begin{subequations}
\begin{align}
  & \partial_T \, G_{a_1 a_2}^{(2)}[0,T] = [\vec{V}_{a_1} (T)]_{i_2} + [\vec{V}_{a_2} (T)]_{i_1} \,, \\
  & \partial_T \vec{V}_{a_1} (T) = M(T) \vec{V}_{a_1} (T) + \mathcal{C}_{a_1} \av{\vec{c}(T)} \,, \\
  & \partial_T \vec{V}_{a_2} (T) = M(T) \vec{V}_{a_2} (T) + \mathcal{C}_{a_2} \av{\vec{c}(T)} \,, \\
  & \partial_T \av{\vec{c}(T)} = M(T) \av{\vec{c}(T)} \,.
\end{align}
\end{subequations}
Note that if $a_1 = a_2 = a$, the first equation reduces to
$\partial_T \, G_{aa}^{(2)}[0,T] = 2 [\vec{V}_{a} (T)]_{i_a}$ and
$\vec{V}_{a_1} = \vec{V}_{a_2} = \vec{V}_{a}$.
Notice as well that $[\vec{V}_{a_i} (T)]_0 = \int_0^T \av{a_i^\dagger a_i (t_1)} dt_1$,
that is the integrated population of the mode $a_i$.
Thereby, the normalized integrated cross-correlation is expressed as
\begin{equation}
    g^{(2)}_{a_1 a_2}[0,T] = \frac{G^{(2)}_{a_1 a_2} [0,T]}{[\vec{V}_{a_1} (T)]_0 [\vec{V}_{a_2} (T)]_0}
    \,.
\end{equation}
Higher correlations can be computed in the same way by deriving the integrated correlator under the 
integral sign. For instance, the $n$-th order integrated correlator of the $a_1$ mode reads
\begin{equation}
G_{a_1 \dots a_1}^{(n)} [0,T] = \int_0^T \dots \int_0^T \mathcal{G}^{(n)}_{a_1 \dots a_1} (t_1, \dots, t_n) dt_1 \dots dt_n =
\int_0^\infty \dots \int_0^\infty \theta(T - t_1) \dots \theta(T - t_n) \mathcal{G}^{(n)}_{a_1 \dots a_1} (t_1, \dots, t_n)
dt_1 \dots dt_n
\,,
\end{equation}
which, after unraveling the equations of motion, leads to the following set of equation
\begin{subequations}
\label{eq:EqsMot_Gn}
\begin{align}
  & \partial_T \, G_{a_1 \dots a_1}^{(n)} [0,T] = n [\vec{V}_{(
a_1)^{n-1}} (T)]_{i_1} \,, \\
  & \partial_T \vec{V}_{(
a_1)^{n-1}} (T) = M(T) \vec{V}_{(a_1)^{n-1}} (T) +
  (n-1) \mathcal{C}_{a_1} \vec{V}_{(a_1)^{n-2}} (T) \,, \\
  & \vdots \\
  & \partial_T \vec{V}_{(a_1)^2} (T) = M(T) \vec{V}_{(a_1)^2} (T) + 2 \,
  \mathcal{C}_{a_1} \vec{V}_{a_1} (T) \,, \\
  & \partial_T \vec{V}_{a_1} (T) = M(T) \vec{V}_{a_1} (T) + 
  \mathcal{C}_{a_1} \av{\vec{c}(T)} \,, \\
  & \partial_T \av{\vec{c}(T)} = M(T) \av{\vec{c}(T)} \,,
\end{align}
\end{subequations}
with the initial conditions $G_{a_1 \dots a_1}^{(n)} [0,0] = \vec{V}_{(a_1)^k} (0) = 0$, for
$k = 1, \dots, n-1$, and where $\vec{V}_{(a_1)^k} (T)$ can be defined in a similar way as Eq.~\eqref{eq:Va1def}
\begin{equation}
\label{eq:Va1ndef}
    \vec{V}_{(a_1)^k} (T) = \int_0^\infty \dots \int_0^\infty \theta(T - t_1) \dots
    \theta(T - t_k) \vec{v}_{a_1 \dots a_1} (t_1, \dots, t_k \,,T) \, dt_1 \dots dt_k \,,
\end{equation}
where
\begin{equation}
    \vec{v}_{a_1 \dots a_1} (t_1, \dots, t_k \,, T) =
    \av{\mathcal{T}_+ [a_1^\dagger (t_1) \dots a_1^\dagger (t_k)] \, \vec{c} (T) \,
    \mathcal{T}_- [a_1 (t_k) \dots a_1 (t_1)]
    } \,.
\end{equation}
It is noteworthy to mention that $G^{(k)}_{a_1 \dots a_1} [0,T] = [\vec{V}_{(a_1)^k} (T)]_0$, meaning that
all the integrated correlations, up to $n$ photons, can be obtained from this set of equations.
\\\\
However, if the time domain is not the same for all the coordinates $t_i$---for example,
when we have two different time bins, Early (E) and Late (L)---the set of equations is not enough
to obtain the correlations of (E,L) or (L,E) time domains. 
The extension of the method is straightforward 
and, in fact, it is formally equivalent to the Quantum Regression Theorem.\\\\
We rewrite the two-time integrated correlation function (Eq.~\eqref{eq:G2Int_gen}) as
\begin{equation}
    G_{a_1 \rightarrow a_2}^{(2)}[0,T; \tau] =
    \int_0^T \int_0^{T+\tau} \, \mathcal{G}^{(2)}_{a_1 \rightarrow a_2} (t_1,t_2) 
    \, dt_1 dt_2 \,,
    =
    \int_0^\infty \int_0^\infty \, \theta(T-t_1)\theta(T+\tau-t_2)
    \theta(t_2-t_1)\mathcal{G}^{(2)}_{a_1 \rightarrow a_2} (t_1,t_2) 
    \, dt_1 dt_2 \,,
\end{equation}
which reduces to $G_{a_1 \rightarrow a_2}^{(2)}[0,T]$ when $\tau = 0$.
We now derive with respect of $\tau$, yielding
\begin{equation}
\partial_\tau G_{a_1 \rightarrow a_2}^{(2)}[0,T; \tau] =
\int_0^\infty \int_0^\infty \, \theta(T-t_1)\delta(T+\tau-t_2)
    \theta(t_2-t_1)\mathcal{G}^{(2)}_{a_1 \rightarrow a_2} (t_1,t_2) 
    \, dt_1 dt_2 =
    \int_0^\infty \theta(T-t_1) \mathcal{G}^{(2)}_{a_1 \rightarrow a_2} (t_1,T+\tau) dt_1 \,.
\end{equation}
Then, in order to compute the previous integral, we need to define the quantity
\begin{equation}
\vec{U}_{a_1} (T,\tau) = \int_0^\infty \theta (T - t_1) \vec{v}_{a_1} (t_1,T +\tau) dt_1 \,,
\end{equation}
and it is easy to check that $\vec{U}_{a_1} (T,\tau = 0) = \vec{V}_{a_1} (T)$. 
Differentiating with respect of $\tau$, and using QRT afterwards, eventually leads to 
\begin{equation}
\partial_\tau \vec{U}_{a_1} (T,\tau) = 
\int_0^\infty \theta (T - t_1) \partial_\tau \vec{v}_{a_1} (t_1,T +\tau) dt_1 =
\int_0^\infty \theta (T - t_1) M(t+\tau) \vec{v}_{a_1} (t_1,T +\tau) dt_1 =
M(T+\tau) \vec{U}_{a_1} (T,\tau) \,.
\end{equation}
Then, we finally obtain the set of equations
\begin{subequations}
\begin{align}
    & \partial_\tau G_{a_1 \rightarrow a_2}^{(2)}[0,T; \tau] = [\vec{U}_{a_1} (T,\tau)]_{i_2} \,, \\
    & \partial_\tau \vec{U}_{a_1} (T,\tau) = M(T+\tau) \vec{U}_{a_1} (T,\tau) \,, 
\end{align}
\end{subequations}
with initial conditions $G_{a_1 \rightarrow a_2}^{(2)}[0,T; \tau = 0] = 
G^{(2)}_{a_1 \rightarrow a_2}[0,T]$ and
$\vec{U}_{a_1} (T,\tau = 0) = \vec{V}_{a_1} (T)$.
\\\\
Therefore, we can express the 2-photon integrated correlation function for the
(E,L) domain as
\begin{multline}
    \int_0^T \int_T^\infty \mathcal{G}^{(2)}_{a_1 a_2} (t_1,t_2) dt_1 dt_2=  \\
    \int_0^T \int_0^\infty \mathcal{G}^{(2)}_{a_1 a_2} (t_1,t_2) dt_1 dt_2 - 
    \int_0^T \int_0^T \mathcal{G}^{(2)}_{a_1 a_2} (t_1,t_2) dt_1 dt_2 =
    G^{(2)}_{a_1 \rightarrow a_2} [0,T; \tau \rightarrow \infty] -
    G^{(2)}_{a_1 a_2} [0,T] \,,
\end{multline}
which requires to compute the correlators $G^{(2)}_{a_1 \rightarrow a_2}[0,T]$ and
$\vec{V}_{a_1} (T)$ first, in the same way one obtains two-time correlations
by solving the one-time observables in first place and subsequently using the QRT to propagate the solutions to get the two-time correlation functions.
Of course, the correlations of reverse process, that is, $a_2 \rightarrow a_1$ is
obtained by swapping $a_1 \leftrightarrow a_2$ in the previous equations.
\section{Photon-counting formula}\label{appendix:PhotCount}

We investigate the time structure of the emission of (at least) two photons, that can be either distinguishable or indistinguishable.
For such a purpose, we set two time bins. The time domain is then split in two: $0 < t \leq T$, that defines the first bin,
and $t > T$, to which we assigned the labels Early (E) and Late (L), respectively. If we extend the description to two sensors and, thereby
two different times $(t_1, t_2)$ (the first associated to the mode/detector $a$, and
the second to $b$), the resulting space is split in four domains: $(\mathrm{I})$, if both photons are detected in the Early bin;
$(\mathrm{II}_a)$, when $a$ is detected in the Late bin and $b$ in the Early one; $(\mathrm{II}_b)$, for the reversed process ($a$ in E and $b$ in L); and
$(\mathrm{III})
$, whether both photons are detected in the Late bin. From the correlations in time $\mathcal{G}_{ab} (t_1, t_2)$ we can observe
the spread in time of the detection plus the asymmetry of the emission. This provides information about the order of the process and
which is its relevant time scale. However, it is difficult to quantify how likely is the emission to occur in a particular way. 
For instance, which is the probability of detecting both photons in $(\mathrm{I})$? To answer this question we make use of the
photon-counting theory~\cite{carmichael_open_1993, gardiner_quantum_2000}.\\\\ The central role is given by the
time-integrated intensity operators
\begin{subequations}
\begin{align}
  \Omega_{a_i,\mathrm{E}} = & \ \xi_i \gamma_i \int_0^{T}
  (a_i^\dagger a_i) (t_1) dt_1  \,,  \\
  \Omega_{a_i,\mathrm{L}} = & \ \xi_i \gamma_i \int_{T}^\infty  (a_i^\dagger a_i) (t_1) dt_1 \,,
\end{align}
\end{subequations}
where $\xi_i$ corresponds to the detectors efficiency
(which, for the sake of simplicity, we assume to be unity
hereafter), $\gamma_i$ are characteristic emission rates
of each mode $a_i$. \\\\
For one mode but two time bins, 
the probability distribution is computed from the (normal-ordered) correlations of these operators $\av{:\Omega_{a,\mathrm{E}}^m \Omega_{a,\mathrm{L}}^n:}$ 
(note that these correlations only converge if the
signal if finite in time)
using the generalized Mandel's formula.
Hence, the probability of detecting in the Early bin 
$m$ and $n$ in the Late bin reads
\begin{equation}
\label{eq:Pmn_formula}
P_{mn} = \frac{1}{m! n! }
\big\langle : \exp(- \Omega_{a}) \, \Omega_{a,\mathrm{E}}^{m} \Omega_{a,\mathrm{L}}^{n}: \big\rangle =
\frac{1}{m! n! } \sum_{k_1,k_2} \frac{(-1)^{k_1+k_2} }{k_1! k_2!}
\big\langle : \Omega_{a,\mathrm{E}}^{m+k_1} \Omega_{a,\mathrm{L}}^{n+k_2}: \big\rangle
\,,
\end{equation}
where $\Omega_a = \Omega_{a, \mathrm{E}} + \Omega_{a, \mathrm{L}}$
is the total time-integrated intensity operator, that is,
\begin{equation}
    \Omega_a = \gamma_a \int_0^\infty dt_1 \ (a^\dagger a)(t_1)
    \,,
\end{equation}
from where we recover the expression for the moments
\begin{equation}
    \av{:\Omega_a^n:} = \gamma_a^n
    \int_0^\infty dt_1 \dots \int_0^\infty dt_n
    \mathcal{G}_a^{(n)} (t_1, \dots, t_n) dt_1 \dots dt_n = \gamma_a^n G_a^{(n)} [0, T \rightarrow \infty] \,,
\end{equation}
from which we can compute the total probabilities, the so-called Mandel's formula,
\begin{equation}
\label{eq:Pn_formula}
P_n = \frac{1}{n!} \av{:\Omega_a^n e^{-\Omega_a}:} = \frac{1}{n!} \sum_k \frac{(-1)^k}{k!} \gamma_a^{n+k} \, G_a^{(n+k)} [0, T \rightarrow \infty] \,,
\end{equation}
which, for practical reasons, we truncate up to $N$ photons, provided that
the probabilities have converged, that is, adding higher-order correlations
do not modify $P_{n \leq N}$. \\\\
For two modes, namely $a$ and $b$, and two time bins, 
the probability distribution is computed from the (normal-ordered) correlations of these operators $\av{:\Omega_{a,\mathrm{E}}^m \Omega_{a,\mathrm{L}}^n
\Omega_{b,\mathrm{E}}^p\Omega_{b,\mathrm{L}}^q:}$ 
(note that these correlations only converge if the
signal is finite in time)
using the generalized Mandel's formula.
Hence, the probability of detecting in the Early bin 
$m_a$ and $m_b$ in the $a$ and $b$ modes, respectively, 
and in the Late bin $n_a$ and $n_b$ reads

\begin{equation}
P_{m_a n_a m_b n_b} = \frac{1}{m_a! n_a! m_b! n_b!}
\big\langle : \exp(- \Omega_a - \Omega_b)\Omega_{a,\mathrm{E}}^{m_a} \Omega_{a,\mathrm{L}}^{n_a}
\Omega_{b,\mathrm{E}}^{m_b}\Omega_{b,\mathrm{L}}^{n_b} : \big\rangle
\,,
\end{equation}
where we have defined the total intensity operators
$\Omega_a = \Omega_{a,\mathrm{E}} + \Omega_{a,\mathrm{L}}$
and $\Omega_b = \Omega_{b,\mathrm{E}} + \Omega_{b,\mathrm{L}}$.
We can rewrite the previous expression after unwinding
the exponential operator
\begin{equation}
P_{m_a n_a m_b n_b} = \frac{1}{m_a! n_a! m_b! n_b!}
\sum_{k_1,k_2,k_3,k_4} \frac{(-1)^{k_1+k_2+k_3+k_4}}{k_1! k_2! k_3! k_4!}
\big\langle : \Omega_{a,\mathrm{E}}^{m_a+ k_1} \Omega_{a,\mathrm{L}}^{n_a + k_2}
\Omega_{b,\mathrm{E}}^{m_b + k_3}\Omega_{b,\mathrm{L}}^{n_b+k_4} : \big\rangle
\,.
\end{equation}
These expressions hardly ever have an analytical solution because, in first place, they require to compute all-order correlations; second, under continuous excitation, the average number of detected photons grows ceaselessly. However, since we are dealing with finite-time excitation, we can assume that the total number of photons obtained during the whole evolution would not exceed a certain number $N$ in most of the cases. In other words, we discard events with more than $N$ photons, as they are highly unlikely to occur, and we can safely take the limit $T \rightarrow \infty$.
Thereby, as a good approximation, we assume
$\av{:\Omega_{a,\mathrm{E}}^m \Omega_{a,\mathrm{L}}^n :} \approx 0$,
for $m+n > N$, or
$\av{:\Omega_{a,\mathrm{E}}^m \Omega_{a,\mathrm{L}}^n
\Omega_{b,\mathrm{E}}^p\Omega_{b,\mathrm{L}}^q:} \approx 0$
for $m+n+p+q > N$. In such cases,
we can approximate the probability distribution 
as a finite sum of these correlations.
For our analysis, truncating the maximum number 
of photons up to 2 is enough in most of the case.
Then, all the probabilities involving more than
two photons, so $m + n > 2$ or $m_a + n_a + m_b +n_b > 2$, are zero.
This is what we call the two-photon approximation (2PA), which is employed to compute the frequency-filtered correlations.
\subsection{One-mode probabilities (two-photon approximation)}
The two-photon events probabilities are
\begin{subequations}
    \begin{align}
        P_{20} = & \ \frac{1}{2} \av{:\Omega_{a,\mathrm{E}}^2:} \,, \\
        P_{11} = & \ \av{:\Omega_{a,\mathrm{E}}\Omega_{a,\mathrm{L}}:} \,, \\
        P_{02} = & \ \frac{1}{2} \av{:\Omega_{a,\mathrm{L}}^2:} \,, 
    \end{align}
\end{subequations}
and the total two-photon probability is computed
using the law of total probability, that is,
adding up all the two-photon contributions, which yields
\begin{equation}
    P_2 = \sum_{m+n = 2} P_{m,n} = P_{20} + P_{11}+ P_{02} =
    \frac{1}{2} \big( \av{:\Omega_{a,\mathrm{E}}^2:} + 2 
    \av{:\Omega_{a,\mathrm{E}}\Omega_{a,\mathrm{L}}:} +
    \av{:\Omega_{a,\mathrm{L}}^2:}\big) = \frac{1}{2}
    \av{:\Omega_a^2:} \,,
\end{equation}
that coincides with the usual definition from Mandel's formula (Eq.~\eqref{eq:Pn_formula}). On the other hand, the one-photon event probabilities are
\begin{subequations}
    \begin{align}
        P_{10} = & \ \av{\Omega_{a,\mathrm{E}}} - 
        \av{:\Omega_{a,\mathrm{E}}^2:} -
        \av{:\Omega_{a,\mathrm{E}} \Omega_{a,\mathrm{L}}:}
        \,, \\
        P_{01} = & \ \av{\Omega_{a,\mathrm{L}}} - 
        \av{:\Omega_{a,\mathrm{L}}^2:} -
        \av{:\Omega_{a,\mathrm{E}} \Omega_{a,\mathrm{L}}:}
        \,.
    \end{align}
\end{subequations}
The total one-photon probability is obtained in the same way
we did for $P_2$
\begin{equation}
   P_1 = P_{10} + P_{01} = \av{\Omega_a} - \av{:\Omega_a^2:} \,.
\end{equation}
Finally, the zero-photon probability is naturally computed
as $P_0 = P_{00} = 1 - P_1 - P_2$. 
\\\\
For instance, for the bare emission of Resonance Fluorescence
($a = \sigma$),
to get the probability distribution, we are required to compute
the following integrated correlations
\begin{subequations}\label{eq:photprobRF}
    \begin{align}
      &  \av{\Omega_{\sigma,\mathrm{E}}} =  \ 
        \gamma_\sigma \int_0^{T} \ n_\sigma (t_1) dt_1  \,, \\
      &  \av{\Omega_{\sigma,\mathrm{L}}} =  \ 
        \gamma_\sigma \int_{T}^\infty \ n_\sigma (t_1) dt_1   = \av{\Omega_\sigma} - \av{\Omega_{\sigma,\mathrm{E}}}
        \,, \\
      & \av{:\Omega_{\sigma,\mathrm{E}}^2:} = \ \gamma_\sigma^2
      \int_0^{T}  \int_0^{T}  \,
      \mathcal{G}^{(2)}_\sigma (t_1,t_2) dt_1 dt_2 \,, \\
      & \av{:\Omega_{\sigma,\mathrm{E}}\Omega_{\sigma,\mathrm{L}}:} = \ \gamma_\sigma^2
      \int_0^{T} \int_{T}^\infty \,
      \mathcal{G}^{(2)}_\sigma (t_1,t_2) dt_1 dt_2 \,, \\
      & \av{:\Omega_{\sigma,\mathrm{L}}^2:} = \ \gamma_\sigma^2
     \int_{T}^\infty \int_{T}^\infty \,
      \mathcal{G}^{(2)}_\sigma (t_1,t_2) dt_1 dt_2 =
      \av{:\Omega_\sigma^2:} - 2 \av{:\Omega_{\sigma,\mathrm{E}}\Omega_{\sigma,\mathrm{L}}:} - \av{:\Omega_{\sigma,\mathrm{E}}^2:}
      \,.      
    \end{align}
\end{subequations}
\subsection{Two-mode probabilities (two-photon approximation)}
For the two-mode case, the non-vanishing probabilities concerning two-photon events are thus
\begin{subequations}
    \begin{align}
    P_{2000} = & \ \frac{1}{2!} \av{:\Omega_{a,\mathrm{E}}^2:} \,, \\
    P_{0200} = & \ \frac{1}{2!} \av{:\Omega_{a,\mathrm{L}}^2:} \,, \\
    P_{0020} = & \ \frac{1}{2!} \av{:\Omega_{b,\mathrm{E}}^2:} \,, \\
    P_{0002} = & \ \frac{1}{2!} \av{:\Omega_{b,\mathrm{L}}^2:} \,, \\
    P_{1100} = & \ \av{:\Omega_{a,\mathrm{E}}
    \Omega_{a,\mathrm{L}}:} \,, \\
    P_{1010} = & \ \av{:\Omega_{a,\mathrm{E}}
    \Omega_{b,\mathrm{E}}:} \,, \\
    P_{1001} = & \ \av{:\Omega_{a,\mathrm{E}}
    \Omega_{b,\mathrm{L}}:} \,, \\
    P_{0110} = & \ \av{:\Omega_{a,\mathrm{L}}
    \Omega_{b,\mathrm{E}}:} \,, \\
    P_{0101} = & \ \av{:\Omega_{a,\mathrm{L}}
    \Omega_{b,\mathrm{L}}:} \,, \\
    P_{0011} = & \ \av{:\Omega_{b,\mathrm{E}}
    \Omega_{b,\mathrm{L}}:} \,.
    \end{align}
\end{subequations}
While the probabilities of the single-photon events are
\begin{subequations}
    \begin{align}
     P_{1000} = & \ \av{\Omega_{a,\mathrm{E}}} - \av{:\Omega_{a,\mathrm{E}}^2:} - \av{:\Omega_{a,\mathrm{E}}
    \Omega_{a,\mathrm{L}}:} - \av{:\Omega_{a,\mathrm{E}}
    \Omega_{b,\mathrm{E}}:} - \av{:\Omega_{a,\mathrm{E}}
    \Omega_{b,\mathrm{L}}:}\,, \\
     P_{0100} = & \ \av{\Omega_{a,\mathrm{L}}} 
     - \av{:\Omega_{a,\mathrm{L}}^2:} - \av{:\Omega_{a,\mathrm{E}}
    \Omega_{a,\mathrm{L}}:} - \av{:\Omega_{a,\mathrm{L}}
    \Omega_{b,\mathrm{E}}:} - \av{:\Omega_{a,\mathrm{L}}
    \Omega_{b,\mathrm{L}}:}\,, \\
     P_{0010} = & \ \av{\Omega_{b,\mathrm{E}}} 
     - \av{:\Omega_{b,\mathrm{E}}^2:} - \av{:\Omega_{a,\mathrm{E}}
    \Omega_{b,\mathrm{E}}:} - \av{:\Omega_{a,\mathrm{L}}
    \Omega_{b,\mathrm{E}}:} - \av{:\Omega_{b,\mathrm{E}}
    \Omega_{b,\mathrm{L}}:}\,, \\
     P_{0001} = & \ \av{\Omega_{b,\mathrm{L}}} 
     - \av{:\Omega_{b,\mathrm{L}}^2:} - \av{:\Omega_{a,\mathrm{E}}
    \Omega_{b,\mathrm{L}}:} - \av{:\Omega_{a,\mathrm{L}}
    \Omega_{b,\mathrm{L}}:} - \av{:\Omega_{b,\mathrm{E}}
    \Omega_{b,\mathrm{L}}:}\,, 
    \end{align}
\end{subequations}
and $P_0 = P_{0000} = 1 - \sum_{m_a + n_a+m_b + n_b \neq 0}
P_{m_a n_a m_b n_b}$. 

Finally, these correlators can be written in terms of 
the integrated correlations are shown above
\begin{subequations}
    \begin{align}
        \av{\Omega_{a_i,\mathrm{E}}} = & \
        \gamma_i \int_0^{T} \ n_{a_i} (t_1) dt_1  \,, \\
        \av{\Omega_{a_i,\mathrm{L}}} = & \
        \gamma_i \int_{T}^\infty \ n_{a_i} (t_1) dt_1  \,, \\
        \av{:\Omega_{a_i,\mathrm{E}} \Omega_{a_j,\mathrm{E}}:} = & \
        \gamma_i \gamma_j \int_0^{T} \int_0^{T}  
        \ \mathcal{G}^{(2)}_{a_i a_j} (t_1,t_2) dt_1  dt_2 \,, \\
        \av{:\Omega_{a_i,\mathrm{E}} \Omega_{a_j,\mathrm{L}}:} = & \
        \gamma_i \gamma_j \int_0^{T} \int_{T}^\infty
        \ \mathcal{G}^{(2)}_{a_i a_j} (t_1,t_2) dt_1  dt_2 \,, \\
        \av{:\Omega_{a_i,\mathrm{L}} \Omega_{a_j,\mathrm{L}}:} = & \
        \gamma_i \gamma_j \int_{T}^\infty \int_{T}^\infty  \
        \mathcal{G}^{(2)}_{a_i a_j} (t_1,t_2) dt_1  dt_2 \,, \\
    \end{align}
\end{subequations}
and the emission rates for the filters (using the sensor method)
have to be $\gamma_i = \gamma_\sigma (\tfrac{\Gamma}{2 \epsilon})^2$. 

%

\section{Spectral correlations for higher Pulse Areas}\label{appendix:HigherPulseAreas}
To ensure completeness in our study, we analyzed frequency-filtered second-order correlations at higher pulse areas ($\Theta = 5\pi, 6\pi$) for both sub-linewidth $(\Gamma = 0.2\gamma_\sigma)$ and linewidth $(\Gamma = 2\gamma_\sigma)$ sensor bandwidths. The increased Rabi rotations manifest in the emission spectrum as new sidebands emerging symmetrically from the central peak, enabling distinct photon correlation regimes across different spectral windows. As shown in Fig. \ref{ap:g2freqHigherPA}, these sidebands introduce broad regions of antibunching—independent of the pulse area—primarily observed in the autocorrelation of the outermost side peaks and their cross-correlation with adjacent sidebands. Conversely, the most pronounced bunching features arise from autocorrelations and cross-correlations of the spectral valleys between the central peak and sidebands. Notably, at even pulse areas (e.g., $\Theta = 6\pi$), cross-correlations between sidebands dominate the antibunching behavior even for broader bandwidths ($\Gamma = 2\gamma_\sigma$). However, narrower antibunching regions observed in the sub-linewidth regime—such as correlations between the central peak shoulders and blue-detuned sidebands—are suppressed when the sensor bandwidth exceeds the spectral resolution of individual features. 

\begin{figure}[h!]
    \centering
    \includegraphics[width = 0.6\columnwidth]{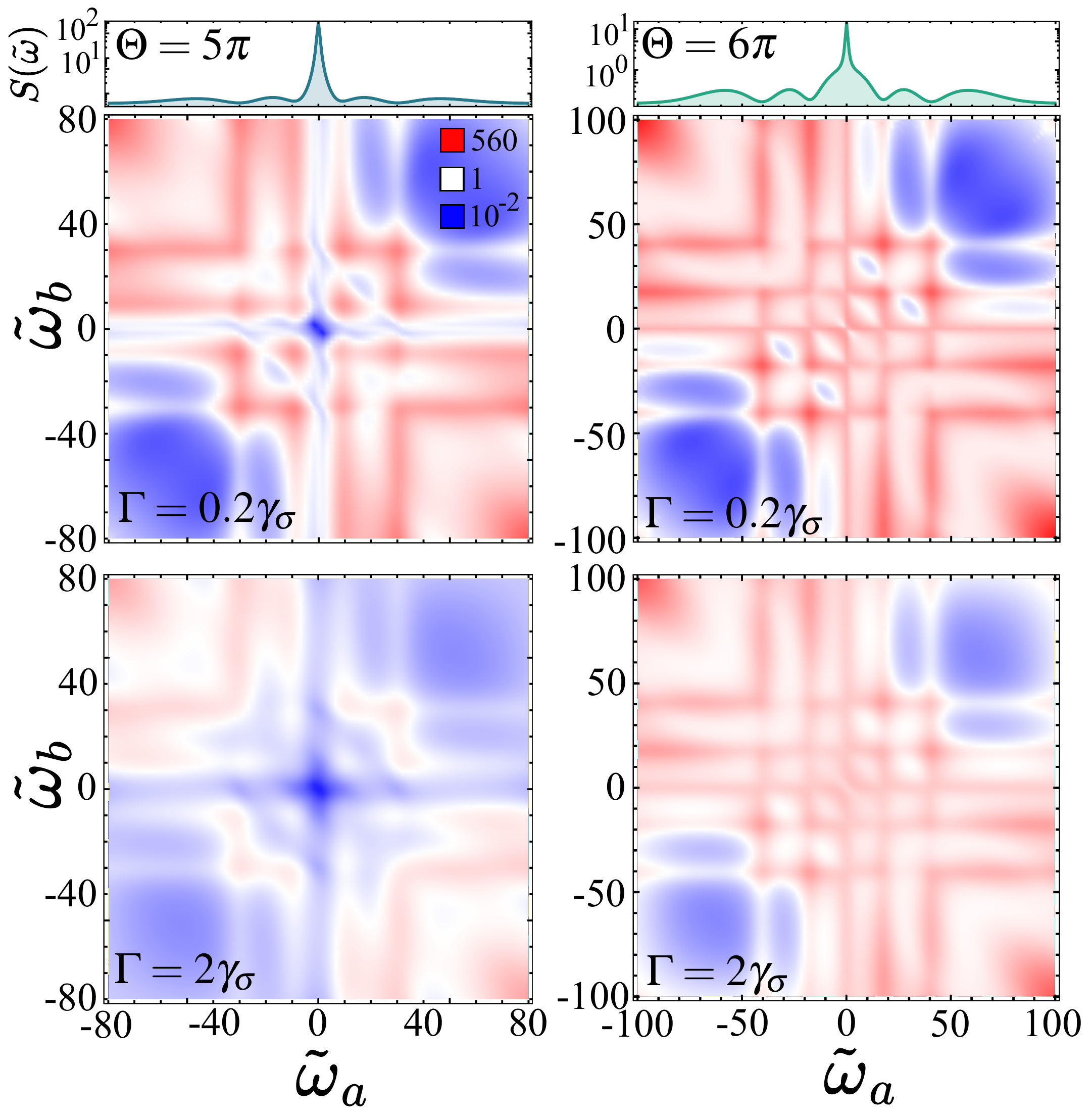}
    \caption{Second order correlation function $g^{(2)}_{ab}[0 ; \Gamma]$ of the detected photons for pulse areas of $5\pi$ (left panel) and $6\pi$ (right panel) for a sensor frequency bandwidth $\Gamma = 2 \gamma_\sigma$.}
    \label{ap:g2freqHigherPA}
\end{figure}

\section{Monte Carlo vs. Two-photon approximation }\label{appendix:MCvsIntegrals}
We compare the filtered time-bin counting probabilities up to 2 photons obtained
from the time-integrated correlations with the exact probabilities extracted from
the Monte Carlo simulations. For narrow filters $\Gamma \lesssim 5 \gamma_\sigma$,
the probabilities (shown in Fig.~\ref{ap:figMCvsInts}), are in good agreement.
However, for wider filters, the results start to differ. The reason is that,
at this point, the contributions from higher photon numbers are not so strongly 
suppressed by the filter. Neglecting these multi-photon contributions drastically 
affects photon-counting statistics, overestimating the two-photon 
probabilities for odd pulse areas and underestimating one-photon ones for
even pulse areas.
\begin{figure}[h!]
    \centering
    \includegraphics[width = 0.8\columnwidth]{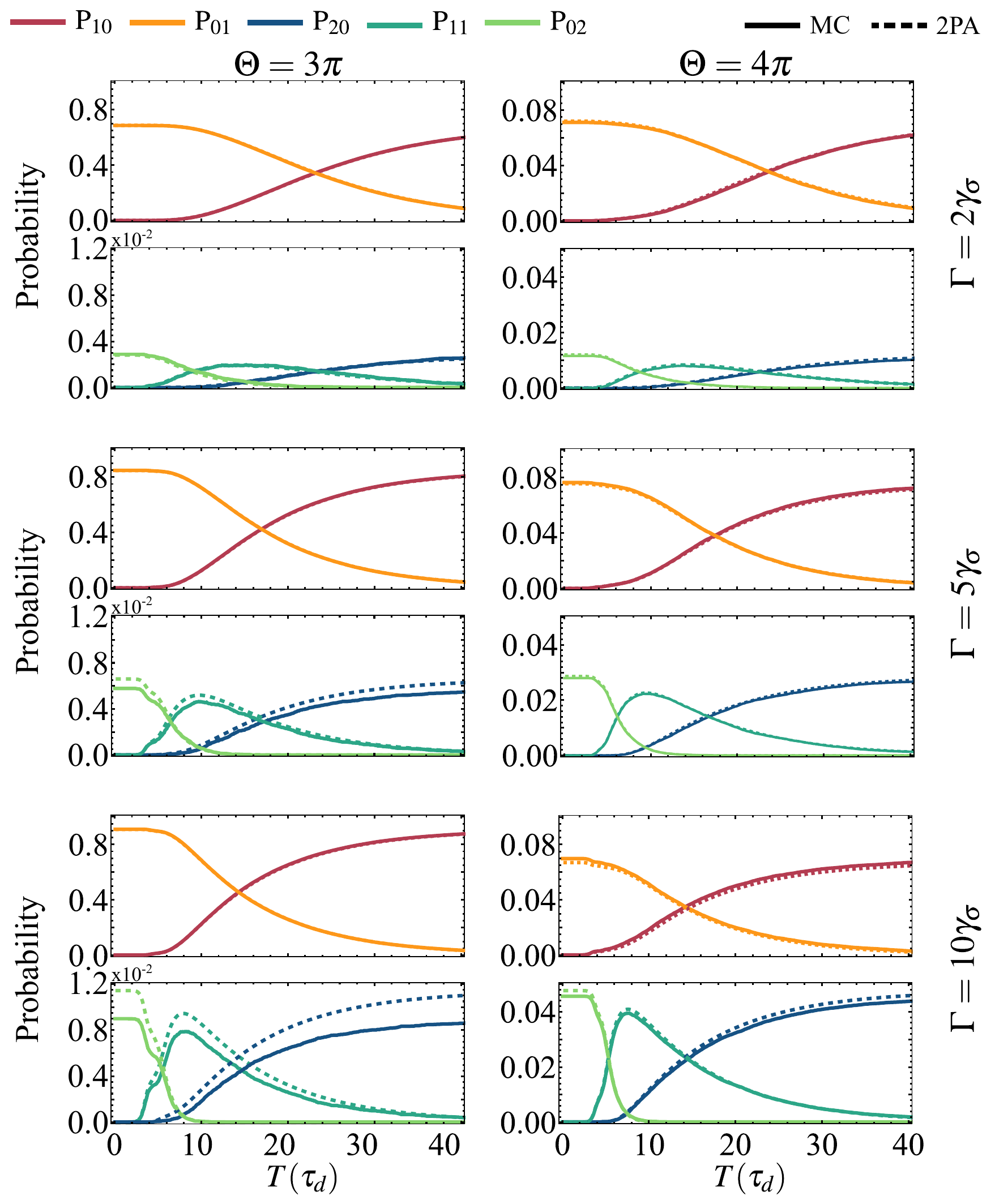}
    \caption{Comparison of the frequency-filtered time-bin probabilities (up to 2 photons) as function of the bin size $T$. The
    full counting statistics from the MC simulations are
    represented by solid lines, whereas the dashed lines denote the probabilities approximated using the integrals up to 2 photons (2PA). The approximation
    hold for the narrower filters $\Gamma = 2, 5 \ \gamma_\sigma$, shown in the first and second rows, 
    respectively. However, it breaks down when the filter
    width is further increased (third row), indicating
    that the higher photon number contributions are
    no longer negligible.}
    \label{ap:figMCvsInts}
\end{figure}

\end{widetext}

\end{document}